\documentclass[a4paper]{article}  
\pdfoutput=1
\usepackage{jheppub}
\usepackage{graphicx,epsfig,wrapfig,amssymb,color,amsmath,bm,breqn}
\usepackage{graphicx}
\usepackage{mathrsfs}
\usepackage{bbm}
\usepackage{indentfirst}
\usepackage{epstopdf}
\usepackage{color}
\usepackage{braket}
\usepackage{placeins}
\usepackage{multirow}
\usepackage{hyperref}
\usepackage{lineno}
\usepackage{slashed}
\usepackage{physics}
\usepackage{comment}
\usepackage{bbold}

\title{Lattice Calculation of the Intrinsic Soft Function and the Collins-Soper Kernel}

\collaboration{\bf{Lattice Parton Collaboration ($\rm {\bf LPC}$)}}

\collaborationImg{\begin{center}\includegraphics[scale=0.1]{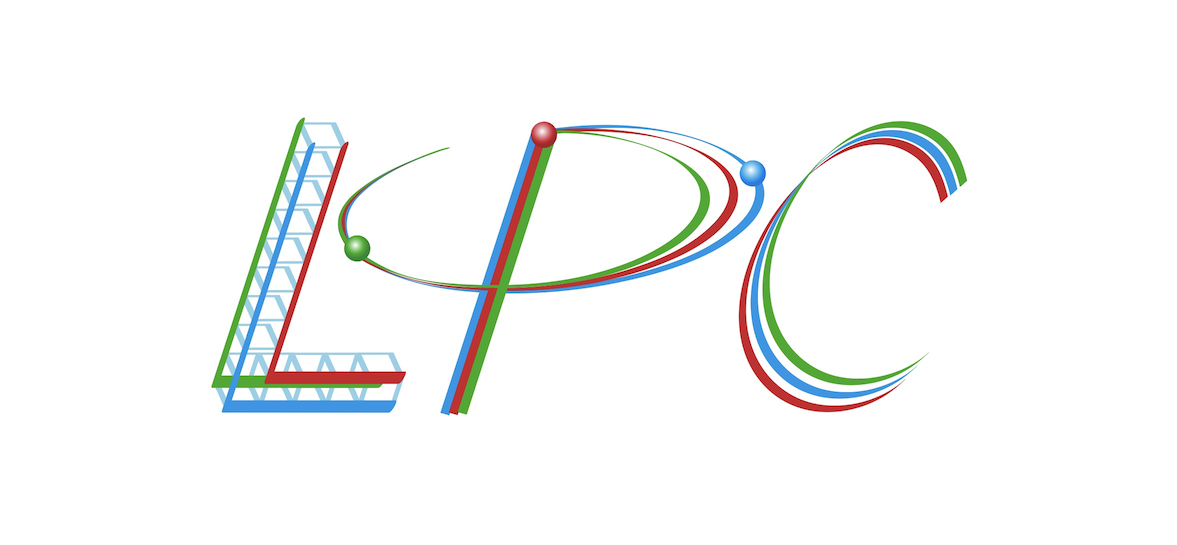}
\end{center}}

\author[ab]{Min-Huan Chu,}
\author[ac]{Jin-Chen He,}
\author[def,\dagger]{Jun Hua,}
\author[def]{Jian Liang,}
\author[c]{Xiangdong Ji,}
\author[g]{Andreas~Sch\"afer,}
\author[g,*]{Hai-Tao Shu,}
\author[c]{Yushan Su,}
\author[g]{Lisa Walter,}
\author[ah]{Wei Wang,}
\author[ij]{Ji-Hao Wang,}
\author[ijkl]{Yi-Bo Yang,}
\author[a]{Jun Zeng}
\author[m]{and Qi-An Zhang}

\affiliation[a]{INPAC, Shanghai Key Laboratory for Particle Physics and Cosmology, Key Laboratory for Particle Astrophysics and Cosmology (MOE), School of Physics and Astronomy, Shanghai Jiao Tong University, Shanghai 200240, China}
\affiliation[b]{Yang Yuanqing Scientiﬁc Computering Center, Tsung-Dao Lee Institute, Shanghai Jiao Tong University, Shanghai 200240, China}
\affiliation[c]{Department of Physics, University of Maryland, College Park, MD 20742, USA}

\affiliation[d]{Key Laboratory of Atomic and Subatomic Structure and Quantum Control (MOE), Institute of Quantum Matter, South China Normal University, Guangzhou 510006, China}
\affiliation[e]{Guangdong Provincial Key Laboratory of Nuclear Science, Institute of Quantum Matter, South China Normal University, Guangzhou 510006, China}
\affiliation[f]{Guangdong-Hong Kong Joint Laboratory of Quantum Matter, Southern Nuclear Science Computing Center, South China Normal University, Guangzhou 510006, China}

\affiliation[g]{Institut f\"ur Theoretische Physik, Universit\"at Regensburg, D-93040 Regensburg, Germany} 

\affiliation[h]{Southern Center for Nuclear-Science Theory (SCNT), Institute of Modern Physics, Chinese Academy of Sciences, Huizhou 516000, Guangdong Province, China}

\affiliation[i]{CAS Key Laboratory of Theoretical Physics, Institute of Theoretical Physics, Chinese Academy of Sciences, Beijing 100190, China}
\affiliation[j]{School of Fundamental Physics and Mathematical Sciences, Hangzhou Institute for Advanced Study, UCAS, Hangzhou 310024, China}
\affiliation[k]{International Centre for Theoretical Physics Asia-Pacific, Beijing/Hangzhou, China}
\affiliation[l]{School of Physical Sciences, University of Chinese Academy of Sciences,
Beijing 100049, China}
\affiliation[m]{School of Physics, Beihang University, Beijing 102206, China}

\emailAdd{$^\dagger$junhua@scnu.edu.cn}
\emailAdd{$^*$hai-tao.shu@ur.de}

\abstract{
We calculate the soft function using lattice QCD in the framework of large momentum effective theory incorporating the one-loop perturbative contributions. The soft function is a crucial ingredient in the lattice determination of light cone objects using transverse-momentum-dependent (TMD) factorization. It consists of a rapidity-independent part called intrinsic soft function and a rapidity-dependent part called Collins-Soper kernel. We have adopted appropriate normalization when constructing the pseudo-scalar meson form factor that is needed in the determination of the intrinsic part and applied Fierz rearrangement to suppress the higher-twist effects. In the calculation of CS kernel we consider a CLS ensemble other than the MILC ensemble used in a previous study. We have also compared the applicability of determining the CS kernel using quasi TMDWFs and quasi TMDPDFs. As an example, the determined soft function is used to obtain the physical TMD wave functions (WFs) of pion and unpolarized iso-vector TMD parton distribution functions (PDFs) of proton.}

\begin{document}
\maketitle

\section{Introduction} 
\label{sec:intro}
The transverse-momentum-dependent parton distribution functions (TMDPDFs) \cite{Collins:1981uk,Collins:1981va,Collins:1984kg}, which encode the probability density for 3D parton momenta in hadrons, have been a topic of intense study in modern hadron physics \cite{Amoroso:2022eow, AbdulKhalek:2022hcn}. The TMDPDFs are universal functions, meaning that they are the same for Drell-Yan (DY) and semi-inclusive deep-inelastic scattering (SIDIS) processes \cite{Scimemi:2019cmh}, up to at most a sign. Both kinds of experiments have been extensively conducted in the past decades, making up our main knowledge for TMDPDFs \cite{Angeles-Martinez:2015sea}. The study of TMDPDFs has a long history, including perturbative, phenomenological and non-perturbative determinations, see \cite{Kang:2022nft, Bury:2020vhj, Bacchetta:2019sam, Echevarria:2020hpy, Bacchetta:2022awv} for a selection of recent publications. TMDPDFs can be obtained from experiments by analyzing the final state particles' transverse momenta phenomenologically \cite{Landry:1999an,Landry:2002ix,Konychev:2005iy,Sun:2014dqm,DAlesio:2014mrz,Echevarria:2014xaa,Kang:2015msa,Bacchetta:2017gcc,Scimemi:2017etj,Bertone:2019nxa,Scimemi:2019cmh,Bacchetta:2019sam,Bury:2022czx, Bacchetta:2022awv}. 
Such fits always require some non-trivial selection of data, see e.g. Fig.~3 and Tab.~3 in \cite{Bury:2022czx} for a recent example. The hard scale $Q$ has to justify the use of perturbation theory and perturbative factorization while, e.g., TMDPDFs are non-perturbative objects depending on $q_{\perp}$ and its Fourier conjugate $b_{\perp}$ respectively. Although in this case there exists a large amount of data, the resulting error bands in Fig.~8 are large (labeled ``ART23"). For other TMDs the experimental data situation is much worse, see, e.g., \cite{Horstmann:2022xkk}. Therefore, combining experimental data with lattice QCD results probably provides the only realistic chance to, e.g., fully determine all eight leading twist TMDs of a nucleon. Such Lattice QCD calculations for TMDs can be grouped in two types. One is based on standard operator product expansion (OPE) plus some additional assumptions to calculate a limited number of Mellin moments of the ratios of TMDPDFs \cite{Hagler:2009mb, Musch:2011er, Yoon:2015ocs, Yoon:2017qzo}. The other follows one of a number of relatively new, more or less equivalent approaches, of which we use the framework of Large Momentum Effective Theory (LaMET) \cite{Ji:2013dva, Ji:2014gla}. In addition to TMDPDF, TMD wave functions (TMDWF) are another important quantity in hadron physics, especially for the description of exclusive reactions. TMDWF provides a description for the partonic structure of hadrons in terms of probability amplitudes. It can be obtained using lattice QCD and LaMET as well \cite{LPC:2022ibr, Chu:2023jia}.

LaMET is based on the observation that parton light cone correlations in the rest frame of the hadron, can be obtained from time-independent spatial correlations in the infinite-momentum frame by continuum perturbation theory. At finite but large hadron momenta, LaMET provides a systematic way to determine TMDs via lattice simulations. To do so the universal soft function, which is the focus of this contribution, plays an important role~\cite{Ji:2014hxa}. LaMET greatly expands the application of lattice QCD in hadron physics, as reviewed, e.g., in \cite{Ji:2020ect}. The soft function accounts for non-cancelling soft gluon-radiation at fixed $Q_{\perp}$ \cite{Collins:1981uk}. It consists of a rapidity-independent part called intrinsic soft function $S_I$ \cite{LatticeParton:2020uhz} and a rapidity evolution kernel called Collins-Soper (CS) kernel $K$ \cite{Collins:1981va}.

The intrinsic soft function was first introduced in \cite{Ji:2019sxk} to eliminate the regulator-scheme-dependence of the quasi TMDPDF/TMDWF. It can be accessed either from heavy quark effective theory \cite{Ji:2019sxk} or via large-momentum-transfer form factors of light mesons \cite{Ji:2019sxk}. The latter possibility has been explored on the lattice using tree level matching \cite{LatticeParton:2020uhz, Li:2021wvl}. The intrinsic soft function was also calculated perturbatively at one-loop order recently in \cite{Deng:2022gzi}. In this work, we will compare the extracted intrinsic soft function using lattice QCD for two different ensembles from the CLS and MILC collaborations. We impose proper normalization and include the one-loop contributions for the first time in a lattice QCD determination. We also apply Fierz rearrangement to suppress higher-twist contaminations \cite{Li:2021wvl}.

The CS kernel can be obtained from global fits of scattering TMDPDFs data collected primarily for DY and SIDIS processes. It can also be extracted from lattice calculable ratios of either Mellin moments of quasi TMDPDFs/beam functions \cite{Ebert:2018gzl,Shanahan:2020zxr, Shanahan:2021tst, Schlemmer:2021aij, Shu:2023cot} or TMD wave functions (WFs)  via a matching procedure. Both tree-level matching \cite{LatticeParton:2020uhz, Li:2021wvl, LPC:2022ibr} and one loop matching \cite{Chu:2023jia} have been explored. In this work, we extract the CS kernel for a CLS ensemble in the framework of LaMET, trying to include one-loop contributions as in \cite{Chu:2023jia}. Besides, we investigate the pros and cons of extracting the CS kernel from quasi TMDWFs and quasi TMDPDFs. We also compare our results with previous ones, based on experimental \cite{Li:2016ctv, Vladimirov:2016dll, Scimemi:2019cmh, Bacchetta:2022awv} and lattice data \cite{ LPC:2022ibr, Shanahan:2021tst, Shu:2023cot}.

With the CS kernel and intrinsic soft function, a lattice determination of physical TMDWFs/TMDPDFs based on the factorization Eq.(\ref{eq:factorization_pdf}) and Eq.(\ref{eq:factorization_wf}) becomes feasible. In \cite{Chu:2023jia} the physical TMDWFs are calculated for the first time and in \cite{LPC:2022zci} the physical unpolarized TMDPDF of the proton is investigated for the first time on a MILC ensemble. In this work,  we discuss TMDWFs and TMDPDFs as applications of the soft function in TMD physics. We estimate the size of discretization uncertainties by comparing the results for the MILC and CLS ensemble. 

The paper is structured as follows: In Sec. \ref{sec:framework} we give the theoretical framework of this work. In Sec. \ref{sec:Si} we provide the details for the calculation of an intrinsic soft function and in Sec. \ref{sec:cskernel} for the CS kernel. We discuss the application of the soft function for TMDWFs and TMDPDFs in Sec. \ref{sec:app}. A summary is given in Sec. \ref{sec:summary}.

\section{Theoretical framework and lattice setup}
\label{sec:framework}

The LaMET factorization formula that relates the quasi TMDPDF $\tilde{f}$ to the light cone TMDPDF $f$ reads \cite{Xiong:2013bka,Ji:2019ewn,Ebert:2022fmh}
\begin{align}
\label{eq:factorization_pdf}
\tilde{f} \left(x, b_{\perp}, \zeta_z, \mu\right) \sqrt{S_I\left(b_{\perp}, \mu\right)}= H_{\Gamma}\left(\frac{\zeta_z}{\mu^2}\right) e^{\frac{1}{2} \ln \left(\frac{\zeta z}{\zeta}\right) K\left(b_{\perp}, \mu\right)}  f\left(x, b_{\perp}, \mu, \zeta\right)  
+\mathcal{O}\big(\frac{\Lambda_{\mathrm{QCD}}^2}{\zeta_z}, \frac{M^2}{\left(P^z\right)^2}, \frac{1}{b_{\perp}^2 \zeta_z}\big),
\end{align}
where $x$ denotes the longitudinal momentum fraction which is the Fourier conjugate to $zP^z$ with $z$ being the offset of the quark-antiquark pair in longitudinal direction and $P^z$ the hadron momentum. $b_{\perp}$ is the transverse separation that is Fourier conjugate to the transverse momentum $Q_{\perp}$. $M$ is the hadron mass and $\zeta$ is a reference rapidity scale that can be chosen at will. The high power corrections are suppressed by the (inverse) hadron momentum and are collected in $\mathcal{O}(...)$. Quasi TMDPDFs are also functions of renormalization scale $\mu$ and $\zeta_z=(2xP^z)^2$. The hard kernel function $H_{\Gamma}=e^h$ is known at next-to-leading order for TMDPDFs \cite{Ji:2019ewn}
\begin{equation}
\begin{aligned}
h^{(1)}=\frac{\alpha_s C_F}{2 \pi}\left(-2+\frac{\pi^2}{12}+\ln\frac{\zeta_z}{\mu^2}-\frac{1}{2}\ln^2\frac{\zeta_z}{\mu^2}\right).
    \label{eq:hard_kernel_pdf}
\end{aligned}
\end{equation}

As found in Ref. \cite{Deng:2022gzi} the matching of TMDWFs is similar to Eq.~(\ref{eq:factorization_pdf}) but slightly modified. 

\begin{eqnarray}
\tilde{\Psi}^{\pm}\left(x, b_{\perp}, \zeta_z, \mu\right) \sqrt{S_I\left(b_{\perp}, \mu\right)}&=& H^{\pm}(x,\zeta_z, \mu) e^{\frac{1}{2} \ln \left(\frac{\mp\zeta_z+i\epsilon}{\zeta}\right) K\left(b_{\perp}, \mu\right)}\Psi^{\pm}\left(x, b_{\perp}, \mu, \zeta\right)  \nonumber\\
& +&\mathcal{O}\big(\frac{\Lambda_{\mathrm{QCD}}^2}{x^2\zeta_z}, \frac{M^2}{\left(xP^z\right)^2}, \frac{1}{x^2b_{\perp}^2 \zeta_z}\big).
\label{eq:factorization_wf}
\end{eqnarray}
Note that following conventions in the literature we use $\zeta_z=(2P^z)^2$ for quasi TMDWFs. The hard kernel function $H^{\pm}=e^{h^{\pm}}$ at next-to-leading order has also a different form \cite{Ji:2020ect, Ji:2021znw, Deng:2022gzi}
\begin{equation}
\begin{aligned}
    h^{\pm(1)}=\frac{\alpha_s C_F}{4 \pi}\left(-\frac{5 \pi^2}{6}-4+l_{\pm}+\bar{l}_{\pm}-\frac{1}{2}\left(l_{\pm}^2+\bar{l}_{\pm}^2\right)\right), \label{eq:hard_kernel_wf}
\end{aligned}
\end{equation}
where $l_{\pm}=\ln[(-x^2 \zeta_z\pm i\epsilon)/\mu^2]$, $\bar{l}_{\pm}=\ln[(-\bar x ^2 \zeta_z\pm i\epsilon)/\mu^2]$, $\bar x = 1-x$. The subscript $\pm$ corresponds to the direction of the Wilson line in quasi TMDWF.

\begin{figure}[thb]
    \centerline{
\includegraphics[width=0.75\textwidth]{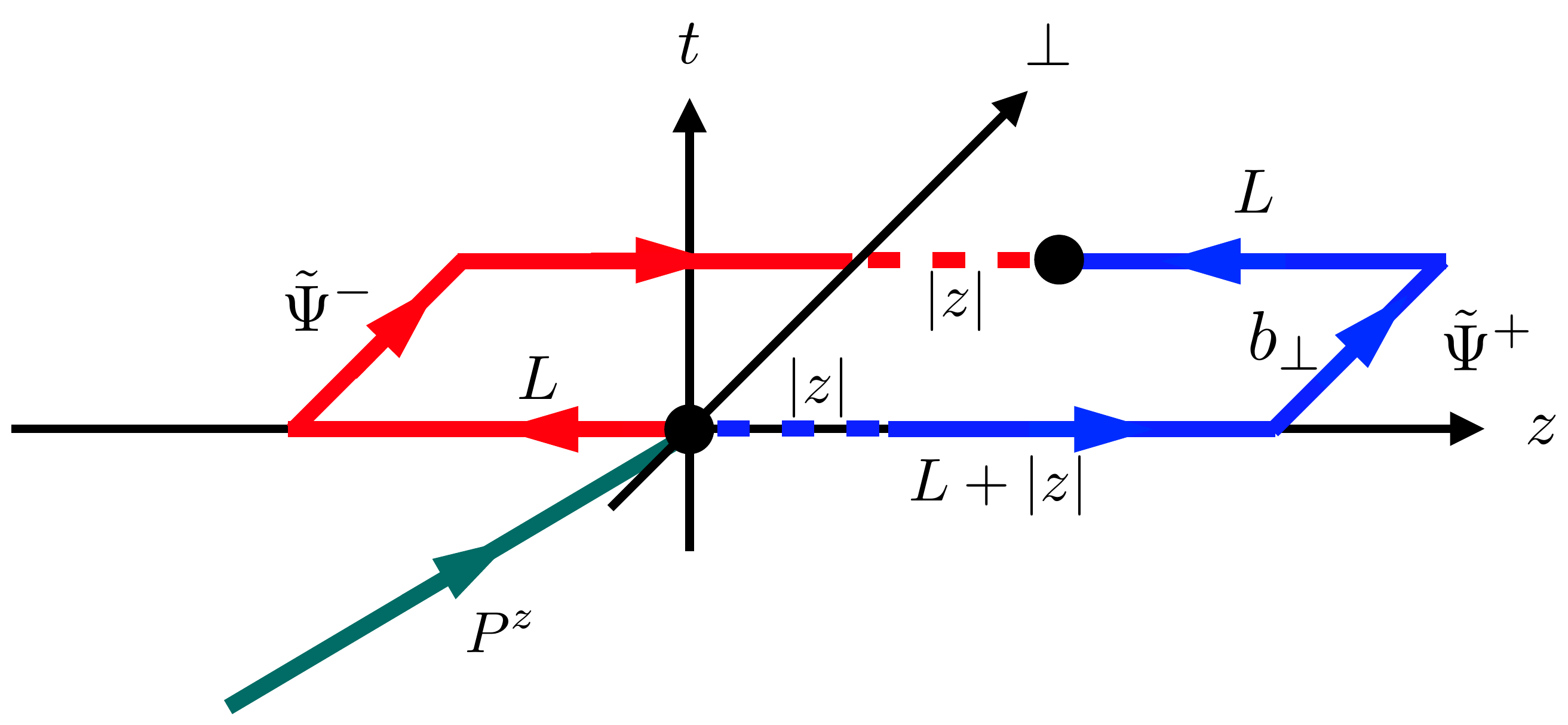}
    }
    \caption{A sketch of the non-local quark-antiquark (black dots) current stretching in $+/-$ direction (assuming the momentum to point into the positive $z$ direction). The offset of the two quarks in $z$-direction is shown as dashed lines of length $z$.}
    \label{fig:sketch}
\end{figure}

The renormalized quasi TMDWF in momentum space is defined as \cite{Zhang:2022xuw}  
\begin{equation}
\tilde{\Psi}^{\pm}\left(x, b_{\perp}, \mu, P^z\right) =\int \frac{P^z d z}{2 \pi} e^{i x z P^z} \lim _{L \rightarrow \infty} \frac{\tilde{\Psi}^{\pm}(L, z,b_{\perp}, P^z)}{\sqrt{Z_E\left(2 L + |z| , b_{\perp}\right) }Z_O(1/a,\mu)},
\label{eq:ren-quasi_TMDWF}
\end{equation}
containing mainly three parts: the bare quasi TMDWF $\tilde{\Psi}^{\pm}(L, z,b_{\perp}, P^z)$ in position space, the Wilson loop $Z_E$ of length $2L+|z|$ and width $b_{\perp}$, and a quark Wilson line vertex renormalization factor $Z_O$. The latter two are part of the renormalization procedure which we will return to in the next section. $a$ denotes the lattice spacing. The bare quasi TMDWF in position space reads  
\begin{equation}
\tilde{\Psi}^\pm(L, z,b_{\perp},P^z) =\frac{\left\langle 0\left|\bar{q}
\left(z \hat{n}_z+b_{\perp} \hat n_\perp\right)
\gamma^t \gamma_5U_{c\pm} q(0)\right| \pi\left(P^z\right)\right\rangle}{\left\langle 0\left|\bar{q}
\left(0 \hat{n}_z+0 \hat n_\perp\right)
\gamma^t \gamma_5 q(0)\right| \pi\left(P^z\right)\right\rangle},
\label{eq:cont-quasi-TMDWF}
 \end{equation}
a correlation function which is constructed by inserting a non-local quark-antiquark ($q\bar{q}$) current between the vacuum and an external pion state $|\pi\rangle$. The current is built by connecting the quark-antiquark pair by a staple-shaped Wilson link of length $L$ (or $L+|z|$ for the longer leg) and width $b_{\perp}$
\begin{align}
U_{c\pm}\equiv 
U_z^\dagger(z \hat{n}_z+b_{\perp} \hat n_\perp; - \bar{L}_\pm) U_\perp(\bar{L}_\pm\hat{n}_z + z\hat{n}_z;b_\perp) U_z(0\hat{n}_z; \bar{L}_\pm + z),
\label{eq:staplewilsonline}
\end{align}
where $U_{\mu}(x;l)\equiv U_{\mu}(x,x+l\hat{n}_{\mu})$ and $\bar{L}_\pm \equiv \pm \mathrm{max}(L,L\mp z)$. Fig.~\ref{fig:sketch} depicts how the non-local current is structured: the quark-antiquark pair is shown as black dots connected by the staple-shaped Wilson line shown as thick blue/red lines stretching in $+/-$ direction. If the longitudinal Wilson line points into the positive direction we put use the ``$+$" sign as superscript in Eq.~(\ref{eq:ren-quasi_TMDWF}), otherwise we put ``$-$". In the folowing we take the latter as an example to illustrate our analysis.

The determination of TMDWF from the quasi ones requires the intrinsic soft function. The intrinsic soft function was first proposed in LaMET to deal with the divergence related to the emission of soft gluons, which is not cancelled by the real and virtual perturbative corrections. Fortunately according to LaMET it can be isolated and turns into an intrinsic function that can be determined non-perturbatively at small transverse momentum using lattice QCD \cite{Ji:2020ect}.  \cite{Ji:2019sxk, Deng:2022gzi} have established a general approach to determine the intrinsic soft function using the quasi TMDWFs $\tilde{\Psi}^{\pm}$ and a pseudo-scalar light-meson form factor of a transversely-separated product of currents~\cite{Ji:2019sxk,Ji:2020ect,Deng:2022gzi}:
\begin{equation}
F\left(b_{\perp}, P_1, P_2,\Gamma,\mu\right) = \frac{\left\langle P_2\left| \bar{q}(b_\perp) \Gamma q(b_\perp)  \bar{q}(0) \Gamma^{\prime} q (0)  \right| P_1\right\rangle}
{\left\langle 0\left|\bar{q}(0) \gamma^\mu \gamma^5 q(0)\right| P_1\right\rangle \left\langle P_2\left|\bar{q}(0) \gamma_\mu \gamma^5 q(0)\right| 0\right\rangle}.
\label{eq:form_factor-cont}
\end{equation}
$P_{1}$ and $P_{2}$ denote momenta in opposite directions along the $z$-axis and are always of equal size in our calculations. The different choices for $\Gamma$ (=$\Gamma^\prime$) project out contributions of different twists, an issue we will address in the next section. The intrinsic soft function then reads \cite{Deng:2022gzi}
\begin{equation}
\begin{aligned}
S_I\left(b_{\perp}, P_1, P_2, \mu\right)=\frac{F\left(b_{\perp}, P_1, P_2,\Gamma,\mu\right) }{\int d x_1 d x_2 H\left(x_1, x_2, \Gamma\right) \tilde{\Psi}^{\pm *}\left(x_2, b_{\perp}, P^z\right) \tilde{\Psi}^{\pm }\left(x_1, b_{\perp}, P^z\right)}, 
\label{eq:soft_def}
\end{aligned}
\end{equation}
where $H\left(x_1, x_2, \Gamma\right)$ is another kernel function known at one loop order \cite{Deng:2022gzi, Ji:2021znw}. For $\Gamma=\mathbb{1}$ or $\Gamma=\gamma_5$ it reads
\begin{equation}
H(x_1,x_2,\Gamma)=H^{(0)} \Big{\{}1+\frac{\alpha_s C_F}{2\pi}\big{[}2+\pi^2+\frac{1}{2}\ln^2(-\frac{x_2}{x_1}\mp i0)+\frac{1}{2}\ln^2(-\frac{\bar{x}_2}{\bar{x}_1}\mp i0)-\ln\frac{16x_1x_2\bar{x}_1\bar{x}_2P^{z4}}{\mu^4}\big{]} \Big{\}},
\end{equation}
while for $\Gamma=\gamma^{\perp}$ or $\Gamma=\gamma^{\perp}\gamma_5$ it reads
\begin{equation}
H(x_1,x_2,\Gamma)=H^{(0)} \Big{\{}1+\frac{\alpha_s C_F}{2\pi}\big{[}\pi^2-4+\frac{1}{2}\ln^2(-\frac{x_2}{x_1}\mp i0)+\frac{1}{2}\ln^2(-\frac{\bar{x}_2}{\bar{x}_1}\mp i0)+\frac{1}{2}\ln\frac{16x_1x_2\bar{x}_1\bar{x}_2P^{z4}}{\mu^4}\big{]} \Big{\}},
\end{equation}
where 
\begin{align}
H^{(0)}=\begin{cases}1/(4N_c), \ \Gamma=\mathbb{1}\\-1/(4N_c), \ \Gamma=\gamma_5,\gamma^{\perp}\text{ or}\ \gamma^{\perp}\gamma_5.\end{cases} 
\end{align}
To extract the intrinsic soft function using lattice QCD, a precise determination of the form factor and a well defined quasi TMDWF are necessary.  Determing the intrinsic soft function with controlled systematics has developed into a pressing task, in order to expand the range of LaMET applications \cite{LatticeParton:2020uhz, Li:2021wvl, Deng:2022gzi}.

The physical TMDWFs evolve with the rapidity scale $\zeta$ satisfying the following renormalization group equation \cite{Collins:1981va, Collins:1981uk}
\begin{equation}
\label{eq_evo}
2\zeta \frac{\rm{d}}{{\rm{d}}\zeta}\ln \Psi(x, b_{\perp}, \mu, \zeta) = K(b_{\perp}, \mu),
\end{equation}
which in turn provides the simplest way to determine the CS kernel $K(b, \mu)$ from physical TMD data. Physical TMDWFs can be transformed into quasi TMDWFs based on Eq.~(\ref{eq:factorization_wf}). Solving the evolution equation along a constant path of $\mu$ allows to fix the CS kernel from the ratio of quasi TMDWFs at different large momenta. The resulting factorization reads  \cite{LPC:2022ibr}
\begin{equation}
K(b_\perp,\mu, x, P_1^z, P_2^z)=\frac{1}{\ln(P_1^z/P_2^z)}\ln\frac{H^\pm(xP_2^z,\mu)\tilde{\Psi}\pm(x,b_\perp,\mu, P_1^z)}{H^\pm(xP_1^z,\mu)\tilde{\Psi}^\pm(x,b_\perp,\mu, P_2^z)},
\label{eq:CS-p-dep}
\end{equation}
which requires a determination of the (renormalized) quasi TMDWFs $\tilde{\Psi}^\pm(x,b_\perp,\mu, P^z)$ on the lattice. Above argument also holds for TMDPDFs, if one simply replaces the (quasi) TMDWFs with the (quasi) TMDPDFs and the accompanying hard kernel function. In fact, the CS kernel is a fundamental nonperturbative function in QCD which describes the interaction of a parton with the QCD vacuum \cite{Vladimirov:2020umg}. It is believed to be independent of all quantum numbers except for the color representation of the probe. At small $b_{\perp}$, the CS kernel can be reliably determined by perturbative or phenomenological calculations. However at large $b_{\perp}$ where the CS kernel becomes non-perturbative, lattice QCD is the only tool that can handle the situation and lattice QCD is essential to relate TMDs at different scales and provides most valuable complementary information compensating lacking experimental data.

With the intrinsic soft function and quasi TMDs at hand, we capture the correct IR physics to all-orders \cite{Ji:2019ewn, Ji:2019sxk} and by a perturbative matching the physical TMDs can be obtained. We will illustrate this matching procedure at the end of this paper with two examples, one for a TMDWF and the other for a TMDPDF.

Before diving into the concrete calculations we would like to provide information on the ensembles used throughout this paper in Tab.~\ref{Tab:setup}. We use four different ensembles in total. The two CLS ensembles are generated using 2+1 flavor dynamical clover fermions and tree-level Symanzik gauge action. X650 has the same parameters as A654 except for its eightfold larger spatial volume. Note that the light quark and strange quark have the same sea quark mass for these two ensembles. The two MILC ensembles are generated using 2+1+1 flavors of highly improved staggered dynamic quarks \cite{Follana:2006rc}. They also only differ by their spacial volume which is eightfold larger for  a12m130. These ensembles are used in different scenarios: on X650 and a12m310 quasi TMDWFs and form factors are calculated; on A654 and a12m130 quasi TMDPDFs are calculated. In all cases the valence quarks are chosen heavier than the sea quarks for the sake of better signals. The difference due to different spatial size and/or valence/sea masses should be minor~\cite{Li:2021wvl,Chu:2023jia}, but will be explicitely investigated in future work. To further improve the signal, hypercubic (HYP) smeared fat links \cite{Hasenfratz:2001hp} have been used for the staple links in all calculations. In addition, the momentum smearing technique \cite{Bali:2016lva} has been used when calculating the quasi TMDPDFs and Coulomb gauge fixed wall source propagators are used when calculating the quasi TMDWFs and form factors. The last column of the table gives the number of measurements, which is equal to the number of the configurations times the number of different sources used for each configuration.

\begin{table}
\footnotesize
\centering
\begin{tabular}{cclccccccc}
\hline
\hline
Ensemble & $a$(fm) & \ \!$N_{\sigma}^3\times\  N_{\tau}$   & $m^{sea}_{\pi}$  & $m^{val}_{\pi}$ & Measure \\
\hline
X650  ~~& 0.098  ~~& $48^3\times$~ \!48  ~~& 333 MeV         ~~& 662 MeV  &  ~ 911$\times$4  ~   \\\hline
A654  ~~& 0.098  ~~& $24^3\times$~ \!48  ~~& 333 MeV         ~~& 662 MeV  &  ~4923$\times$20  ~   \\\hline
\multirow{2}{*}{a12m130 ~} & \multirow{2}{*}{0.121 ~} & \multirow{2}{*}{$48^3\times$~ \!64 ~~} & \multirow{2}{*}{132 MeV ~} & 310 MeV & 1000$\times$4 \\
& & & & 220 MeV & 1000$\times$16\\ 
\hline
a12m310  ~~& 0.121  ~~& $24^3\times$~ \!64  ~~& 305 MeV         ~~& 670 MeV  &  ~ 1053$\times$8  ~   \\
\hline
\hline
\end{tabular}
\caption{The lattice setups used in this work. X650 and A654 were generated using 2+1 flavors of dynamical clover fermions and tree-level Symanzik gauge action by the CLS collaboration. We remark that for these two ensembles the light quark and strange quark have the same mass in the sea. a12m310 and a12m130 are generated using 2+1+1 flavors of highly improved staggered dynamic quarks \cite{Follana:2006rc} (HISQ) by MILC collaboration \cite{MILC:2012znn}. }
\label{Tab:setup}
\end{table}

\section{Intrinsic soft function}
\label{sec:Si}
As the rapidity independent part of the off-light-cone soft function, the intrinsic soft function $S_I$ eliminates the regulator scheme dependence of the quasi TMDPDF/TMDWF. Its determination relies on the calculation of the quasi TMDWF which we present below.

\subsection{Quasi TMDWF}

\emph{Bare quasi TMDWF.---} From Eq.~(\ref{eq:soft_def}) we know that the first piece needed for the intrinsic soft function is the quasi TMDWF. In this section we show how it is determined, taking X650 ($a=0.098$ fm) as an example. Similar results have been obtained for a12m130 \cite{LPC:2022ibr} and a12m310 \cite{Chu:2023jia} both with valence pion mass 670 MeV. To obtain the bare quasi-TMDWF in position space on the lattice, one first calculates the following two-point correlation
\begin{equation}
C_2^{-}(L,z,b_{\perp},t,P^z)=\sum_{\vec{x}}e^{-{\rm i}P^z \vec{x}\cdot \hat{n}_z}\langle O_{\Gamma}(L,z,b_{\perp},t)O^{\dagger}_{\pi}(0,P^z)\rangle,
\label{eq:barequasiTMDWF}
\end{equation}
where the interpolators read
\begin{equation}
\begin{split}
O_{\Gamma}(L,z,b_{\perp},t)&\equiv \bar{u}(\vec{x}+b_{\perp}\hat{n}_{\perp}+z\hat{n}_z,t)U_{c-}\Gamma d(\vec{x},t)\\
O^{\dagger}_{\pi}(P^z,t)&\equiv \sum_{\vec{x},\vec{y}} \bar{u}(\vec{x},t)\gamma_5 d(\vec{y},t)e^{-i P^z\vec{y}\cdot \hat{n}_z}.
\end{split}
\end{equation}

\begin{figure}[htb]
\centering
\includegraphics[width=0.5\textwidth]{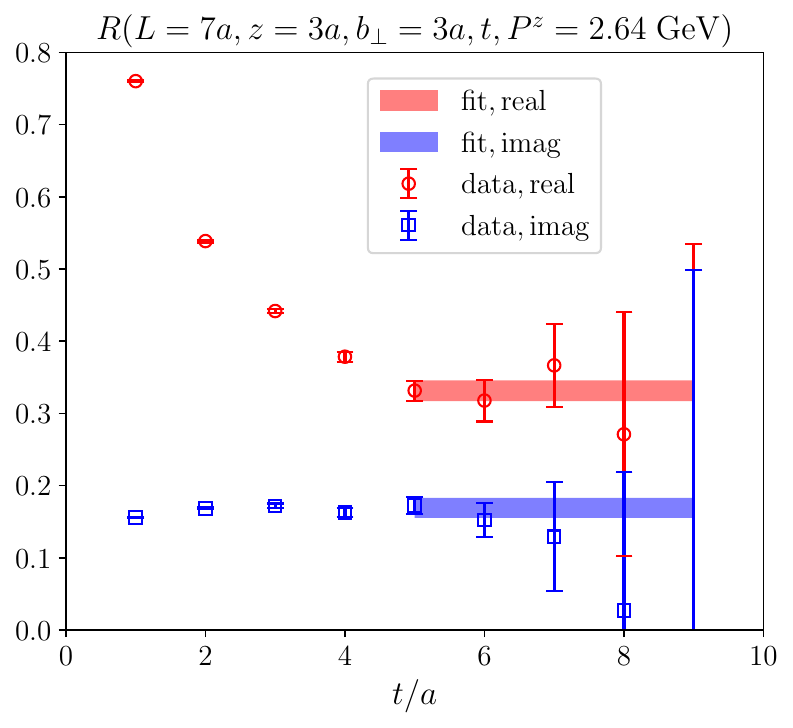}
\caption{Example for a one-state fit in $t$ with momentum $P^z=2.64$ GeV, $b_{\perp}=0.3$ fm, $L=0.7$ fm and $z=0.3$ fm.}
\label{fig:large_t_limit}
\end{figure}

 Ideally one should use $\Gamma=(\gamma^z\gamma_5+\gamma^t\gamma_5)/2$ to eliminate power corrections. However in \cite{LPC:2022ibr} it was demonstrated that the corrections are at most 5\%, such that for simplicity we can just take $\Gamma=\gamma^t\gamma_5$. In this calculation we have $0\leq L \leq 10a$, $0\leq z \leq 10a$, $0 \leq b_{\perp} \leq 7a$, $0\leq t \leq 9a$ and $P^z=\{0, 6, 8, 10, 12\}\times 2\pi/(48a)=\{0, 1.58, 2.11, 2.64, 3.16\}\ \mathrm{GeV}$. To ensure that artifacts are small for the considered momenta we examine the dispersion relation in Appendix \ref{app:dispersion} for the ensemble X650 and a12m310, on which the soft function will be calculated. We also point out that the previous calculation on a12m130 \cite{LPC:2022ibr} only considered $L=7a$, which should suffice as shall be seen later. We normalize the above non-local two-point function with the corresponding local two-point function 
\begin{equation}
\frac{C^{-}_2(L,z,b_{\perp},t,P^z)}{C^{-}_2(L,z=0,b_{\perp}=0,t,P^z)}=\tilde{\Psi}^{-}(L, z,b_{\perp},P^z)\frac{1+c_0(z,b_{\perp},P^z,L)e^{-\Delta Et}}{1+c_1e^{-\Delta Et}},
\label{eq:c2_fit}
\end{equation}
and find that the ground-state contribution, which reproduces the continuum definition Eq~(\ref{eq:cont-quasi-TMDWF}), can be obtained by a one- or two-state fit. In \cite{LPC:2022ibr, Chu:2023jia} both fitting Ansätze are explored and a one-state ansatz was adopted in the end for a better control on the systematic uncertainty in the fits. It is also done here for the same reason. See Fig. \ref{fig:large_t_limit} for an example for such a fit, where we have defined 
\begin{equation}
R(L,z,b_{\perp},t,P^z)\equiv \frac{C_2^-(L,z,b_{\perp},t,P^z)}{C_2^-(L,0,0,t,P^z)\sqrt{Z_E(2L+|z|,b_{\perp})}}
\end{equation}
for simplicity. The figure shows that the one-state Ansatz does capture the correct behavior of the lattice data. More examples are given in Appendix \ref{app:t-fit}.

\begin{figure}[htb]
\centering
\includegraphics[width=0.5\textwidth]{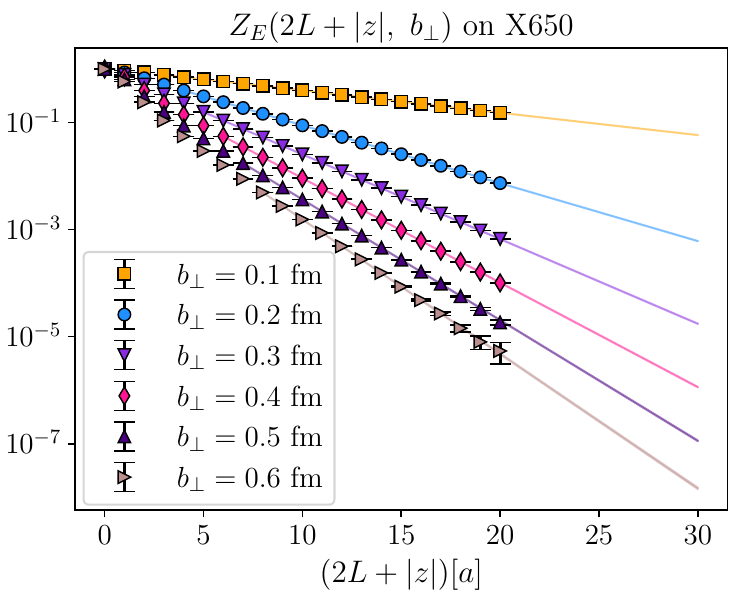}
\caption{Extrapolation of the Wilson loop $Z_E\left(2L+|z| , b_{\perp} \right)$ at $b_{\perp}=\{1,2,3,4,5,6\}a$.}
\label{fig:wiloop_fit_x650}
\end{figure}
\emph{Renormalization.---} 
The bare quasi TMDWFs contain three divergences, the linear divergence originates from the self-energy corrections of the Wilson line, the pinch-pole singularity is caused by the interaction between two legs of the staple-shaped Wilson link and the logarithm divergence is generated by vertices involving Wilson line and light quark. These singularities can be regulated in the way proposed in \cite{Zhang:2022xuw} given by the second line of Eq.~(\ref{eq:ren-quasi_TMDWF}). The square root of the Wilson loop $\sqrt{Z_E(2L+|z|,b_\perp,a)}$ renormalizes the former two singularities \cite{Ji:2017oey, Ishikawa:2017faj, Green:2017xeu, Shanahan:2019zcq, Ji:2020brr} and $Z_O(1/a,\mu)$ renormalizes the last one  \cite{LatticePartonCollaborationLPC:2021xdx,Ji:2021uvr,Zhang:2022xuw}. In Fig.~\ref{fig:wiloop_fit_x650} we show the Wilson loop calculated on X650 and its extrapolation to large $2L+|z|$, where lattice data is either unavailable or too noisy. The extrapolation is feasible because the linear divergence induced by self-energy corrections and gluon exchange effects are exponentially in $2L+|z|$~\cite{Ji:2017oey} and thus can be separated from the rest. The points in the figure denote lattice data and the solid lines are the extrapolated results via one-state fits in the range where we have precise data. We ignore the uncertainties in the extrapolated results as they are negligible compared to other uncertainties, e.g. the statistical uncertainties in the two-point functions. The curves in Fig.~\ref{fig:wiloop_fit_x650} actually contain errors, but these are too small to be visible.

\begin{figure}[htb]
\centering
\includegraphics[width=0.5\textwidth]{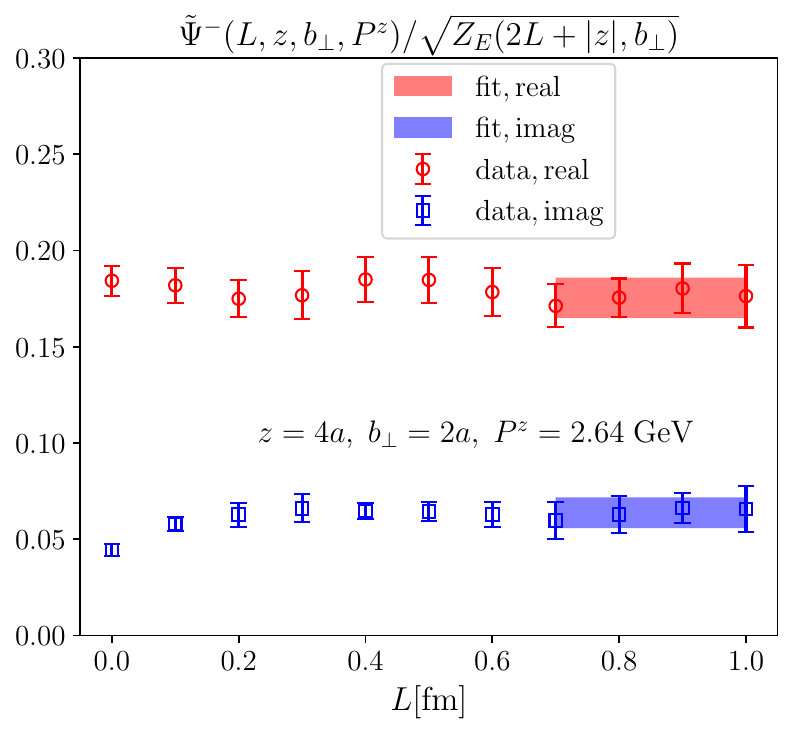}
\caption{Fit to a constant at large $L$  for momentum $P^z=2.64$ GeV, $b_{\perp}=0.2$ fm and $z=0.4$ fm.}
\label{fig:large_L_limit}
\end{figure}

Arguably the large $L$ limit in Eq.~(\ref{eq:ren-quasi_TMDWF}) can be achieved by looking for a plateau in $L$. Inspired by the discovery in Ref.~\cite{Zhang:2022xuw} the ratio
\begin{equation}
\tilde{\Psi}^{-}(z,b_{\perp},\mu,P^z)=\lim _{L \rightarrow \infty}\frac{\tilde{\Psi}^{-}(L, z,b_{\perp},\mu,P^z)}{\sqrt{Z_E\left(2 L + |z| , b_{\perp}\right) }}
\label{eq:large-L}
\end{equation}
is expected to saturate to a constant at $L\simeq 0.7$ fm. This does turn out to be the case for our data, see Fig.~\ref{fig:large_L_limit}, where a plateau can be identified in the range $L\geq 7a=0.7$ fm for a randomly choosen momentum, $b_{\perp}$ and $z$. In fact, we notice that in this figure the plateau appears already at $L\sim 0.4$ fm. As is shown in Appendix \ref{app:L-fit} the plateau appears at larger $L$ for smaller $z$, and $L\simeq 0.7$ fm is always safe, even at $z=0$. 

\begin{figure}[thb]
\centering
\includegraphics[width=0.5\textwidth]{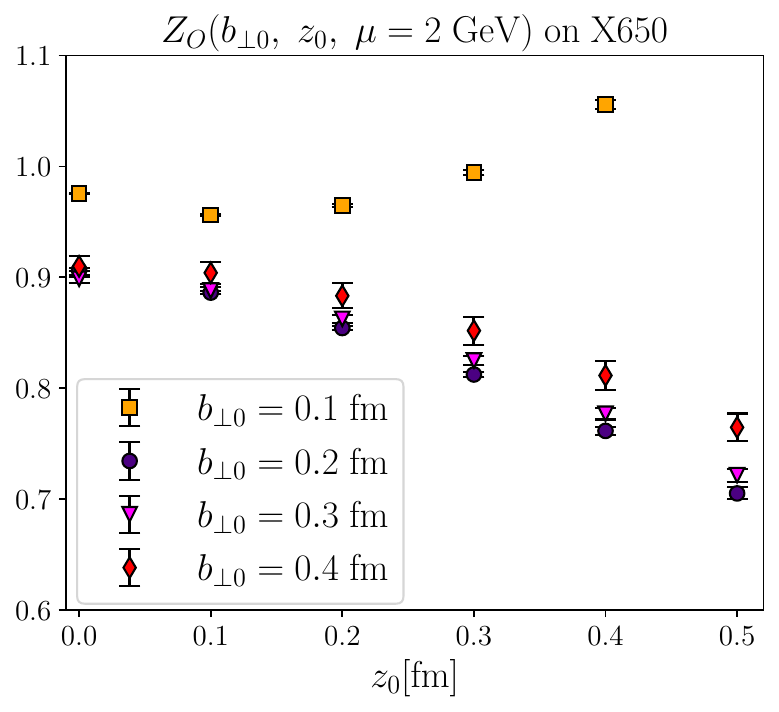}
\caption{The renormalization factor $Z_O$ (see Eq.(\ref{eq:zo_psi_definition})) measured for several $b_{\perp0}$.}
\label{fig:Zo_cls_X650}
\end{figure}

 $Z_O(1/a,\mu)$ can be obtained by taking the ratio of the bare quasi TMDWF calculated at rest on the lattice to the one calculated in the $\overline{\rm MS}$ scheme 
\begin{align}
    Z_O(1 / a, \mu)=\frac{\tilde{\Psi}^{-}\left(z_0, b_{\perp 0}, P^z=0\right)}{\tilde{\psi}^{\overline{\mathrm{MS}}}\left( z_0, b_{\perp 0}, \mu\right)},
    \label{eq:zo_psi_definition}
\end{align}
where \cite{Zhang:2022xuw}
\begin{align}
\label{eq:hMSbar}
\tilde{\Psi}^{\mathrm{\overline{MS}}}\left(z_0, b_{\perp 0}, \mu\right)=1+\frac{\alpha_s C_F}{2\pi}\bigg\{\frac{1}{2}+3\gamma_E-3\text{ln}2+\frac{3}{2}\text{ln}[\mu^2(b_{\perp 0}^2+z_0^2)]-2\frac{z_0}{b_{\perp 0}}\text{arctan}\frac{z_0}{b_{\perp 0}}\bigg\}+{\cal O}(\alpha_s^2).
\end{align}
The $\overline{\rm MS}$ scale is set to $\mu$=2 GeV. $z_0$ and $b_{\perp 0}$ should be fixed to appropriate values where both discretization artifacts and higher twist contaminations get strongly suppressed, indicated by a plateau in $b_{\perp 0}$ observed at some $z_0$, to guarantee the validity of matching to the perturbative calculations. To this aim we calculate $Z_O(1/a, \mu)$ for different $z_0$ and $b_{\perp 0}$, as shown in Fig.~\ref{fig:Zo_cls_X650}. It can bee seen that the best plateau in $b_{\perp 0}$ appears at small $z_0$, starting from $b_0=2a=0.2$ fm. So we choose $z_0=0$ and average the $Z_O$ measured at $b_{\perp 0}=2a$ and $b_{\perp 0}=3a$, which results in $Z_O=0.903(2)$. After dividing by $Z_O$ we add the superscript ``$r$" to the quasi TMDWF to indicate that it has been renormalized by $Z_O$.

%%%%%%%%%%%%%%%%%%%%%%%%%%%%%%%%%%%%%%%%%%%%%%%%%%
%%%%%%%%%%%%%%%%%%%%%%%%%%%%%%%%%%%%%%%%%%%%%%%%%%
\begin{figure}[htb]
\centerline{
\includegraphics[width=0.5\textwidth]{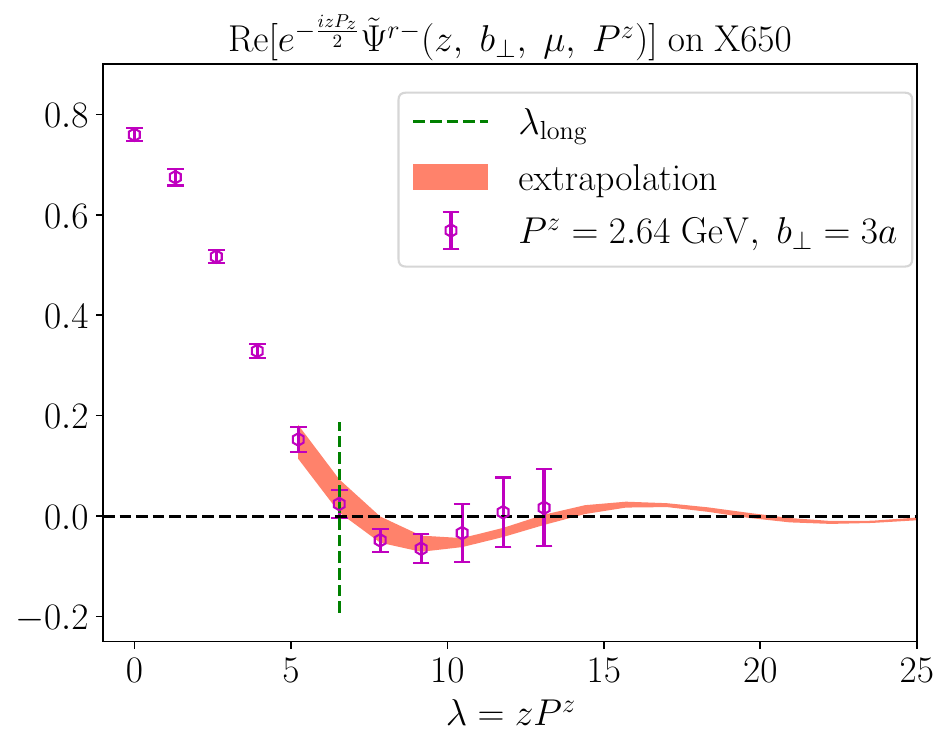}
}
    \caption{Real part of the renormalized quasi TMDWFs (points) and extrapolations for the tail (band) in the large $\lambda$ region. }
    \label{fig:lambda-extrap-x650}
\end{figure}

\begin{figure*}[htb]
\centerline{
\includegraphics[width=0.5\textwidth]{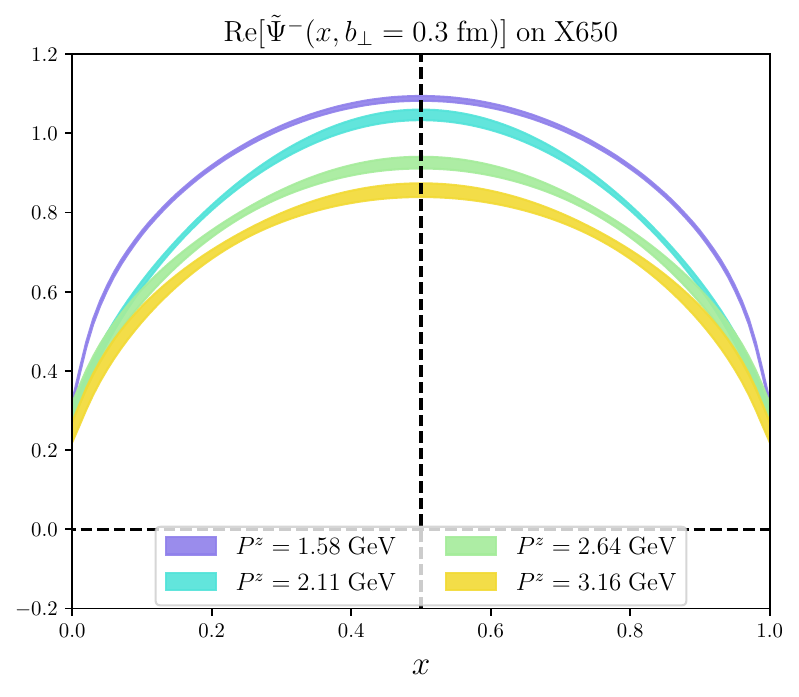}
\includegraphics[width=0.5\textwidth]{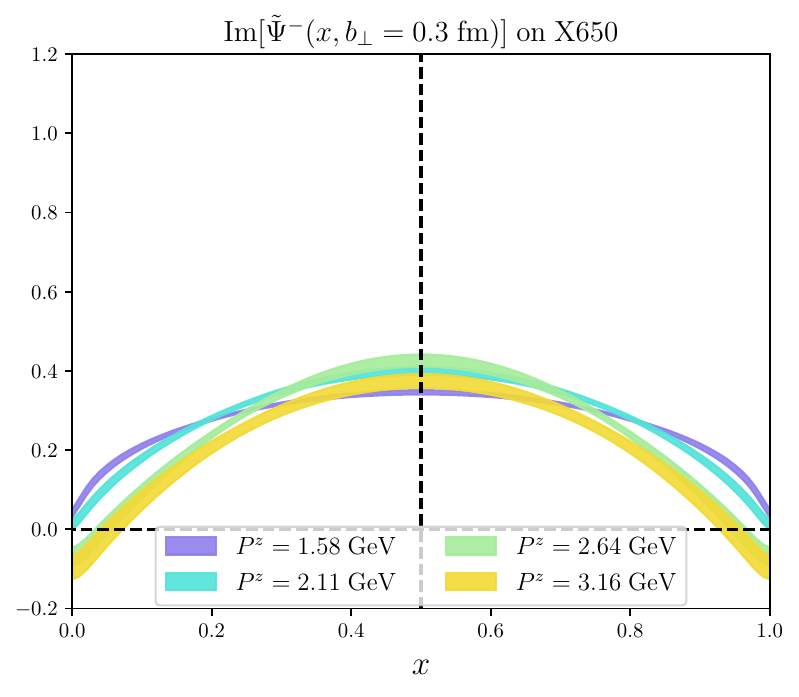}
}
    \caption{The real (left) and imaginary (right) part of the quasi TMDWFs obtained for X650 in momentum space with different $P^z$ for a selected $b_{\perp}$.}
    \label{fig:tmdwf-mom-x650}
\end{figure*}

\emph{Large $\lambda$ extrapolation.---} The Fourier transformation in Eq.~(\ref{eq:ren-quasi_TMDWF}) gets contributions from all real $z$, including very large ones for which lattice simulations are not possible. For this reason an extrapolation to large $z$ (or equivalently $\lambda$) is essential. In Fig.~\ref{fig:lambda-extrap-x650}  we show the real part of the renormalized quasi TMDWFs at available $\lambda=zP^z$ for a selected momentum $P^z=2.64$ GeV as an example. We can see that the quasi TMDWFs approach zero at large $\lambda$ (with larger errors though), which indicates a good convergence when transforming to momentum space. At still larger $\lambda$ we have to extrapolate. We do so, using the following complex ansatz  \cite{Ji:2020brr}
\begin{align}
    \tilde{\Psi}^{r-}_{\mathrm{extra}}(\lambda) = \left[\frac{m_1}{(-i \lambda)^{n_1}}+e^{i \lambda} \frac{m_2}{(i \lambda)^{n_2}}\right] e^{-\lambda / \lambda_0},
    \label{eq:lambda-ansatz}
\end{align}
where $m_1$, $m_2$, $n_1$, $n_2$ are fit parameters. We perform a joint fit of the tails for $b_{\perp}$ in the ranges of $\lambda$ adjusted for different momenta. In the fits $n_1$ and $n_2$ are equal and independent of $b_{\perp}$. $m_1$ and $m_2$ are complex valued different for different $b_{\perp}$. $\lambda_0$ has been set to a large number, safely larger than the possible correlation length at any finite momentum considered here. A detailed discussion of each term in this ansatz can be found in \cite{Ji:2020brr, LPC:2022zci}. We shift the fit ranges by $1a$ and re-perform the fits. The differences of the central values between the two fits are taken as systematical uncertainties. After $\lambda$-extrapolation we Fourier-transform the quasi TMDWFs using
\begin{align}
\tilde{\Psi}^{-}\left(x, b_{\perp},  \mu, P^z\right) = \int\frac{\mathrm{d}\lambda}{2\pi} e^{-ix\lambda} \tilde{\Psi}^{r-}\left(z,b_{\perp},\mu, P^z\right).
\end{align}
The obtained quasi TMDWFs in momentum space are shown in Fig.~\ref{fig:tmdwf-mom-x650} for a moderate $b_{\perp}$ and will be used in the next sections for further calculations.

\subsection{Pseudo-scalar Meson Form Factor}

\begin{figure}[htb]
\centerline{
\includegraphics[width=0.5\textwidth]{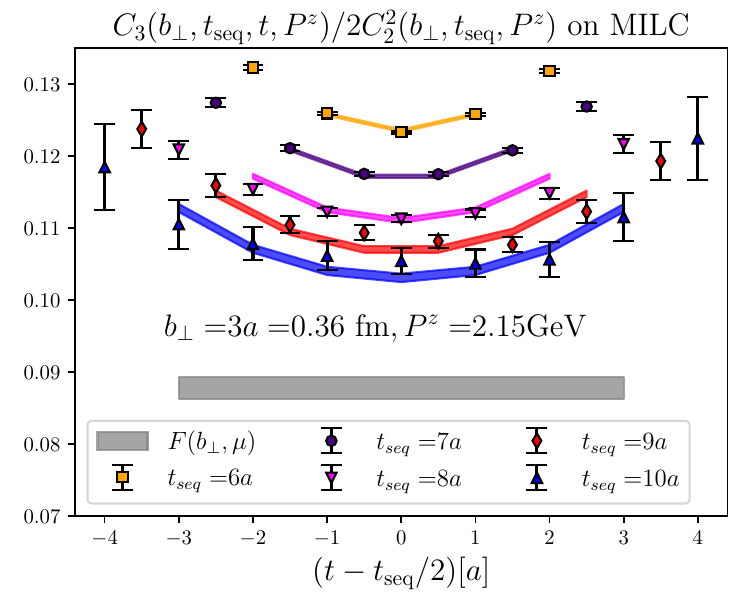}
\includegraphics[width=0.5\textwidth]{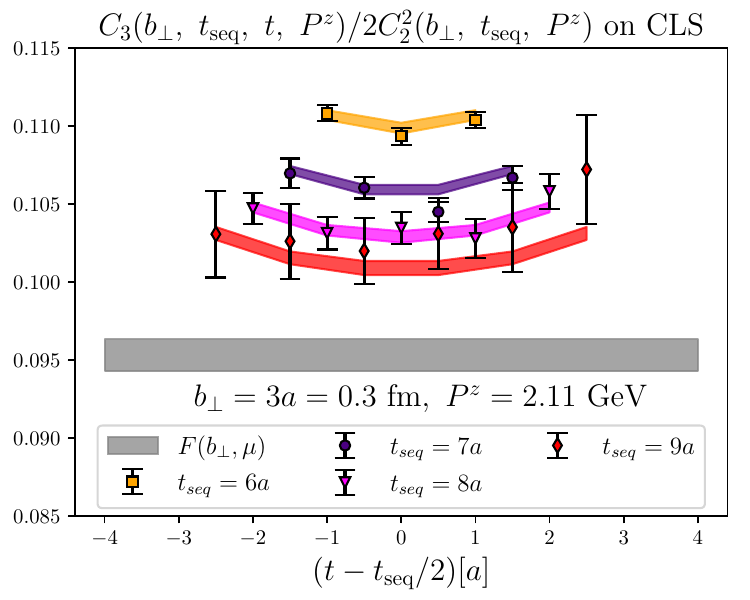}
}
\caption{Extraction of the bare form factor via a joint fit for the ratio Eq.~(\ref{eq:form-factor-lattice}) and the local two-point function Eq.~(\ref{eq:barequasiTMDWF}). The data points are the lattice data of the ratio and the colorful bands denote the fit results. The ground-state contributions (namely the bare form factors) are shown as grey bands. The left panel is for ensemble a12m310 (MILC) and the right panel is for ensemble X650 (CLS).}
\label{fig:c3_c2_fit}
\end{figure}

\emph{Extraction of form factor.---} Another piece appearing in the definition of the intrinsic soft function is the light pseudo-scalar meson form factor. In this section we calculate this form factor for  the two ensembles a12m310 and X650. We choose a12m310 instead of a12m130 based on the practical consideration of computation costs. But we have confirmed that the sea-quark mass effects are marginal, see Appendix \ref{app:sea-quark-mass}. To allow for large momentum extrapolation we consider the three hadron momenta $P^z=\{4, 5, 6\}\times 2\pi/(24a)=\{1.72, 2.15, 2.58\}\ \mathrm{GeV}$ for a12m310 ($a=0.121$ fm) and $P^z=\{6, 8, 10\}\times 2\pi/(48a)=\{1.58, 2.11, 2.64\}\ \mathrm{GeV}$ for X650 ($a=0.098$ fm). We have tried including a fourth, higher momentum and found that the difference is negligible due to the larger errors for the fourth momentum. To obtain the bare form factor on the lattice, one needs to calculate a three-point function 
\begin{equation}
C_3(b_{\perp},\Gamma,\mu,t_{\mathrm{seq}},t,P^z)=\sum_{z^ \prime} e^{-i2z_z^ \prime P^z} \langle O_{\pi}(t_{\mathrm{seq}},-P^z)
\bar{u}\Gamma u(\vec{z}^{\; \prime}+b_\perp, t)
\bar{d}\Gamma d(\vec{z}^{\; \prime},t) O^{\dagger}_{\pi}(0,P^z)\rangle
\end{equation} 
and divide it by the squared local two-point function (let $L=z=b_{\perp}=0$ in Eq.~(\ref{eq:barequasiTMDWF})). Here $t_{\mathrm{seq}}$ is the source-sink separation (source at origin) which is set to $t_{\mathrm{seq}}=\{6,7,8,9\}a$ on X650 and $t_{\mathrm{seq}}=\{6,7,8,9,10\}a$ on a12m310. The interpolators $\bar{u}\Gamma u$ and $\bar{d}\Gamma d$ are inserted at time slice $t$. 
It can be shown that after inserting single particle intermediate states and taking the large (imaginary) time limit ($0\ll t \ll t_{\mathrm{seq}}$), this ratio reproduces the continuum definition Eq.~(\ref{eq:form_factor-cont}). In practice this requires a two-state fit of the following ratio
\begin{equation}
\frac{C_3(b_{\perp},\Gamma,t_{\mathrm{seq}},t,P^z)}{2\left[C_2(0,0,0,t_{\mathrm{seq}}/2,P^z)\right]^2} = F(b_{\perp},\Gamma,P^z)\frac{1+c_1(e^{-\Delta Et}+e^{-\Delta E(t_{\mathrm{seq}}-t)})}{1+c_2e^{-\Delta E t_{\mathrm{seq}}/2}},
\label{eq:form-factor-lattice}
\end{equation} 
 in which the ground state contribution gives the bare form factor $F(b_{\perp},\Gamma,P^z)$. To extract the ground-state contribution one can perform a correlated joint fit for different $t_{\mathrm{seq}}$ of the ratio and the two-point function, which share the excitation energy $\Delta E$. This fit is repeated for every $P^z$ and $b_{\perp}$. We show an example of such a fit for both X650 and a12m310 at a randomly chosen momentum $P^z$ and $b_{\perp}$ in Fig.~\ref{fig:c3_c2_fit}. In the fits the first two and last two data points have been discarded due to their strong excited-state contamination. A closer look at the impact of small $t_{\mathrm{seq}}$ data set to the fit is given in Appendix \ref{app:small_tseq}. 
 For the three-point function we have taken the sum $\sum_{\mu={1,2}} C_3(\Gamma=\gamma^\mu)+C_3(\Gamma=\gamma_5\gamma^\mu)$ to suppress higher-twist effects, which will be discussed in detail in the next paragraph. The extracted bare form factors are renormalized using constants taken from \cite{Bali:2020lwx} for X650 and for a12m310 we find $Z_V/Z_A=0.94(1)$, $Z_S/Z_A=1.11(3)$ and $Z_P/Z_A=0.95(3)$.

\begin{figure}[htb]
\centerline
{
\includegraphics[width=0.5\textwidth]{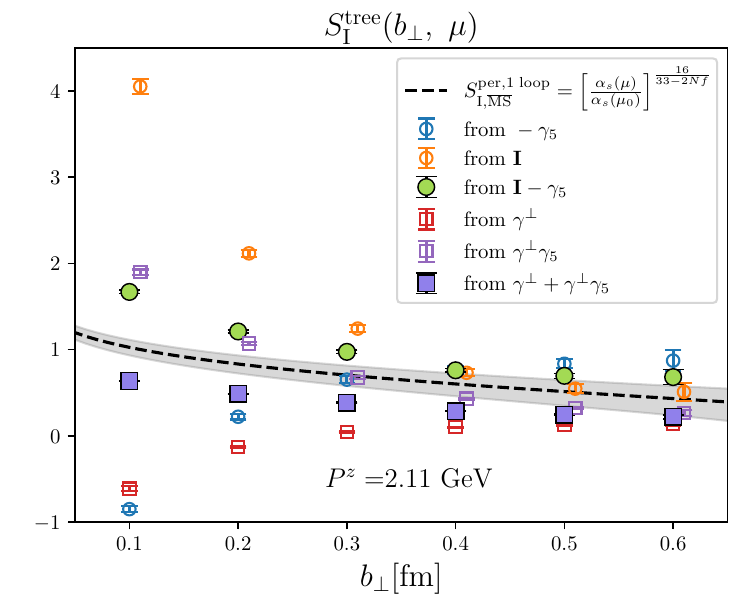}
\includegraphics[width=0.5\textwidth]{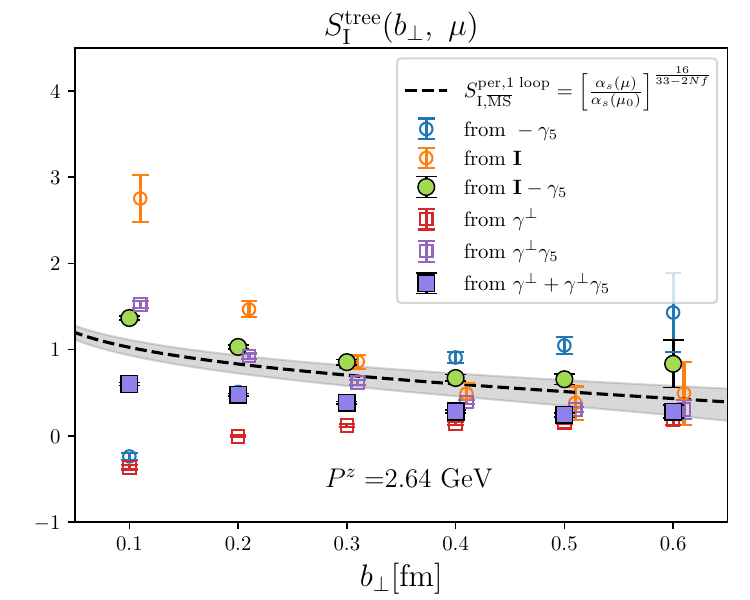}
}
\caption{The intrinsic soft function calculated on X650 using tree-level matching from different channels and two combinations of them. The left panel shows results obtained at 2.11 GeV while the right panel shows results obtained at 2.64 GeV. The data points have been shifted horizontally for better visibility. Only statistical uncertainties are shown.}
\label{fig:soft_function_combine_mom}
\end{figure}

\emph{Fierz rearrangement.---} Inspired by Ref.\cite{Li:2021wvl} one can Fierz-rearrange the four-quark operators to suppress the higher-twist contaminations. There are a few possibilities to do so. For instance one finds that the combination
\begin{equation}
\begin{split}
\sum_{\mu} \left[F(\Gamma=\gamma^{\mu})+F(\Gamma=\gamma^{\mu}\gamma_5)\right]
&=(\bar{\psi}_a\gamma^{x,y}\psi_b)(\bar{\psi}_c\gamma_{x,y}\psi_d)+(\bar{\psi}_a\gamma^{x,y}\gamma_5\psi_b)(\bar{\psi}_c\gamma_{x,y}\gamma_5\psi_d)\\
&=\bar{\psi}_c\gamma^{\mu}\gamma_5\psi_b\bar{\psi}_a\gamma_{\mu}\gamma_5\psi_d+\bar{\psi}_c\gamma^{\mu}\psi_b\bar{\psi}_a\gamma_{\mu}\psi_d
\end{split}
\label{eq:Fierz_perp}
\end{equation}
is dominated by the leading-twist contribution $\gamma^{\mu}\gamma_5$, as the second term in the last line $\bar{\psi}_c\gamma^{\mu}\psi_b\bar{\psi}_a\gamma_{\mu}\psi_d$ vanishes for a pion state. In the second line we have used the property that $F(\gamma^t)$, $F(\gamma^z)$, $F(\gamma^t\gamma_5)$ and $F(\gamma^z\gamma_5)$ vanish for the pion form factor \cite{Deng:2022gzi}. This verifies the advantage in using such a combination of the Dirac structures $F(\gamma^{\perp})$ and $F(\gamma^{\perp}\gamma_5)$ on the lattice to identify the leading-twist contribution. 

\begin{figure*}[htb]
\centerline
{
\includegraphics[width=0.5\textwidth]{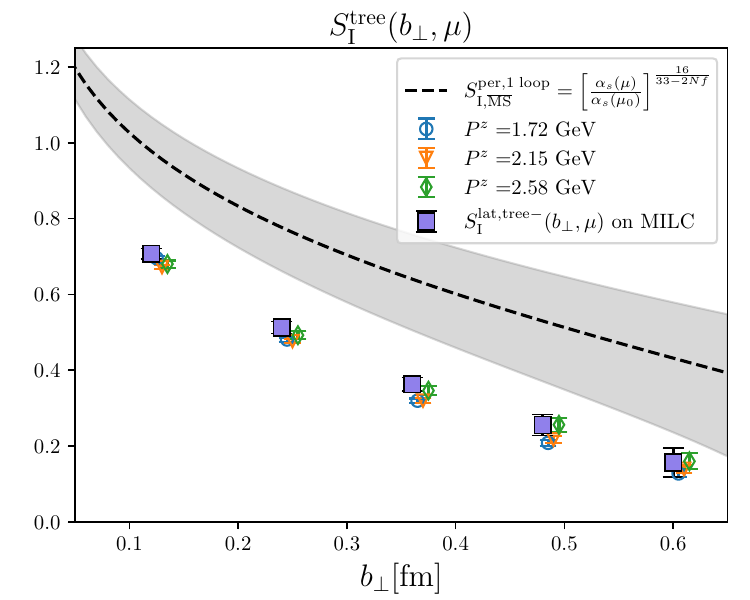}
\includegraphics[width=0.5\textwidth]{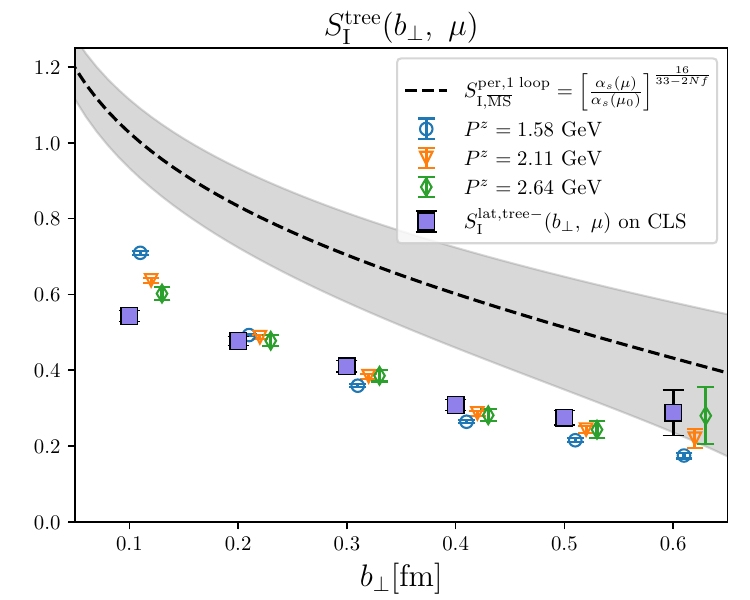}
}
\centerline
{
\includegraphics[width=0.5\textwidth]{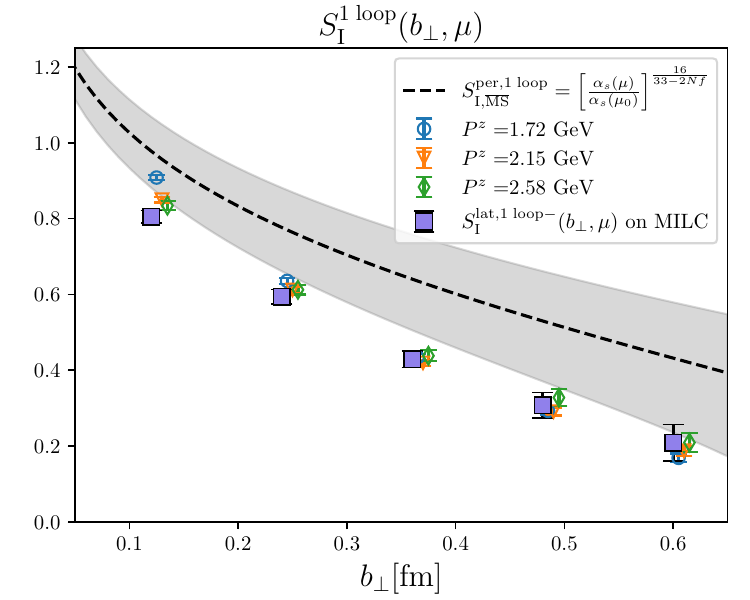}
\includegraphics[width=0.5\textwidth]{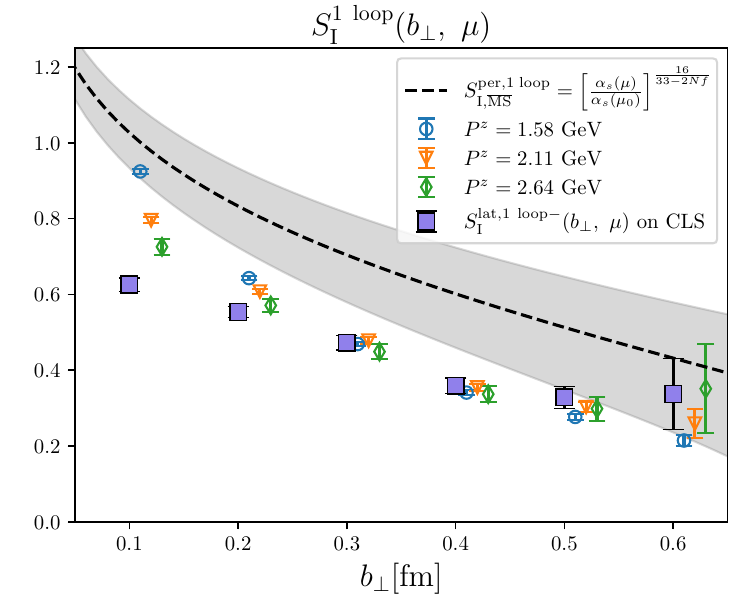}
}
\caption{The intrinsic soft functions obtained from the combination of $F(\gamma^{\perp})$ and $F(\gamma^{\perp}\gamma_5)$ (see Eq.~(\ref{eq:Fierz_perp})). The left panels are from ensemble a12m310 (MILC) while the right panels are from ensemble X650 (CLS). The first row shows results using tree-level matching while the second row is for 1-loop matching. The ``$-$" and ``$+$" sign in the legends indicates the direction of the quasi TMDWF used in the calculation. The data points have been shifted horizontally for better visibility. Only statistical uncertainties are shown.}
\label{fig:soft_function}
\end{figure*}

Similarly the combination of $F(\mathbb{1})$ and $F(\gamma_5)$ 
\begin{equation}
\begin{split}
F(\Gamma=\mathbb{1})-F(\Gamma=\gamma_5)&=(\bar{\psi}_a\psi_b)(\bar{\psi}_c\psi_d)-(\bar{\psi}_a\gamma_5\psi_b)(\bar{\psi}_c\gamma_5\psi_d)\\
&=\frac{1}{2}\bar{\psi}_c\gamma^{\mu}\gamma_5\psi_b\bar{\psi}_a\gamma_{\mu}\gamma_5\psi_d-\frac{1}{2}\bar{\psi}_c\gamma^{\mu}\psi_b\bar{\psi}_a\gamma_{\mu}\psi_d,
\end{split}
\label{eq:Fierz_I5}
\end{equation} 
also gives access to the leading-twist contribution. However, $F(\mathbb{1})$ and $F(\gamma_5)$ have an additional UV divergence \cite{Deng:2022gzi}, leading to a strong momentum dependence, such that this combination is less practical to use, see Fig.~\ref{fig:soft_function_combine_mom}, where we show the intrinsic soft function obtained from form factors calculated on X650 using tree-level matching (let $H(x_1, x_2, \Gamma)=1$ in Eq.~(\ref{eq:soft_def})) at two different momenta from all four channels alone, as well as two combinations of different channels mentioned above. Also shown is the 1-loop perturbative result calculated following \cite{Deng:2022gzi} using the renormalizaton group equation. The error band is determined in the way described in \cite{Chu:2023jia}. From the figure we can see that the intrinsic soft functions from different single channels show strong variation, and even carry opposite sign, especially at small $b_{\perp}$, which is however consistent with the observation in  \cite{Li:2021wvl}. When the momentum increases from 2.11 GeV to 2.64 GeV, a better convergence can be seen at larger momentum, confirming the need for large momentum to eliminate power corrections. Another observation is that the intrinsic soft functions from the two combinations show much better consistency, demonstrating that the higher twist effects can be significantly reduced by the Fierz rearrangement. Comparing the intrinsic soft functions from $\mathbb{1}-\gamma_5$ and $\gamma^{\perp}+\gamma^{\perp}\gamma_5$, it can be seen that the latter shows only a mild dependence on $P^z$ due to the absence of the additional UV divergence \cite{Deng:2022gzi}. Therefore, we use this one  in the following calculation.

\begin{figure*}[htb]
\centerline
{
\includegraphics[width=0.75\textwidth]{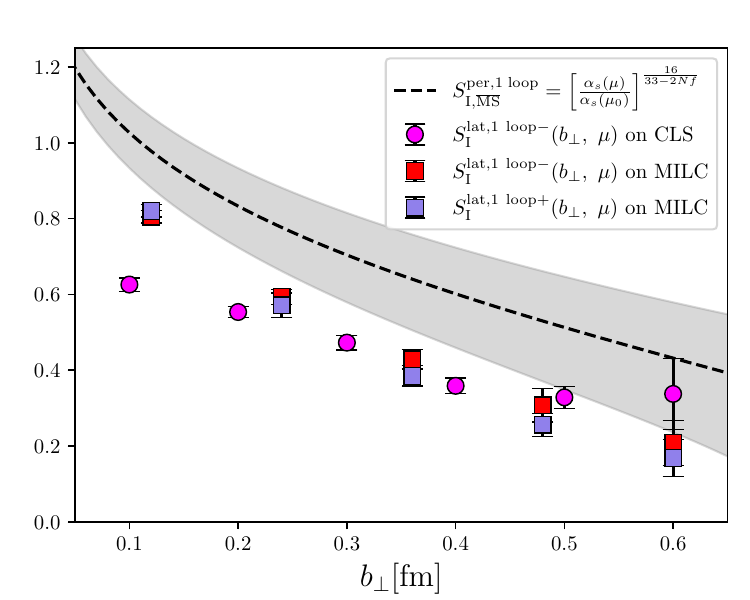}
}
\caption{Final results for the intrinsic soft functions obtained from the CLS and MILC ensembles. $S^{\mathrm{lat,1\;loop}\pm}$ corresponds to the lattice results extracted by $\tilde{\Psi}^{\pm}$.}
\label{fig:final_soft_function}
\end{figure*}

\subsection{Results}

In Fig.~\ref{fig:soft_function} we show the intrinsic soft functions calculated on a12m310 and X650 with tree-level matching and 1-loop matching. Note that the infinite momentum limit is reached only by 
extrapolation using
\begin{equation}
S_{\mathrm{I}}(b_{\perp}, \mu) = S_{\mathrm{I}}(b_{\perp}, \mu, P^z) + \frac{c}{\left(P^z\right)^2}.
\label{eq:mom-extrap-soft}
\end{equation}
In all cases the intrinsic soft functions obtained for X650 show stronger $P^z$-dependence than those for a12m310. When going from tree-level matching to 1-loop matching, the intrinsic soft functions increase significantly for both ensembles, approaching the 1-loop perturbative values, especially at small $b_{\perp}$. Based on all these studies, we regard the results from 1-loop matching and  $\gamma^{\perp}+\gamma^{\perp}\gamma_5$ combination as our final estimates of the intrinsic soft function, summarized in Fig.~\ref{fig:final_soft_function}. Generally speaking, the final intrinsic soft function on two ensembles show satisfactory agreement except at small $b_{\perp}$ where lattice discretization effects are the most significant.

\section{Collins-Soper Kernel}
\label{sec:cskernel}

The CS kernel describes the rapidity evolution of TMDWFs and TMDPDFs. Results containing one-loop contributions were already calculated for a12m130 using quasi TMDWFs in the framework of LaMET in \cite{LPC:2022ibr} and were revisited in \cite{Chu:2023jia} on a12m310. In this section we provide the results for X650 obtained in the same way. Here we use quasi TMDWF in ``$-$" direction and use the 1-loop determination of $H$ \cite{Ji:2020ect, Deng:2022gzi}. usinging the quasi TMDWFs obtained above for different momenta $\{P_1^z, P_2^z\}$ in Eq.~(\ref{eq:CS-p-dep}) we get a momentum-dependent CS kernel. 

\begin{figure}[htb]
\centering
\includegraphics[width=0.5\textwidth]{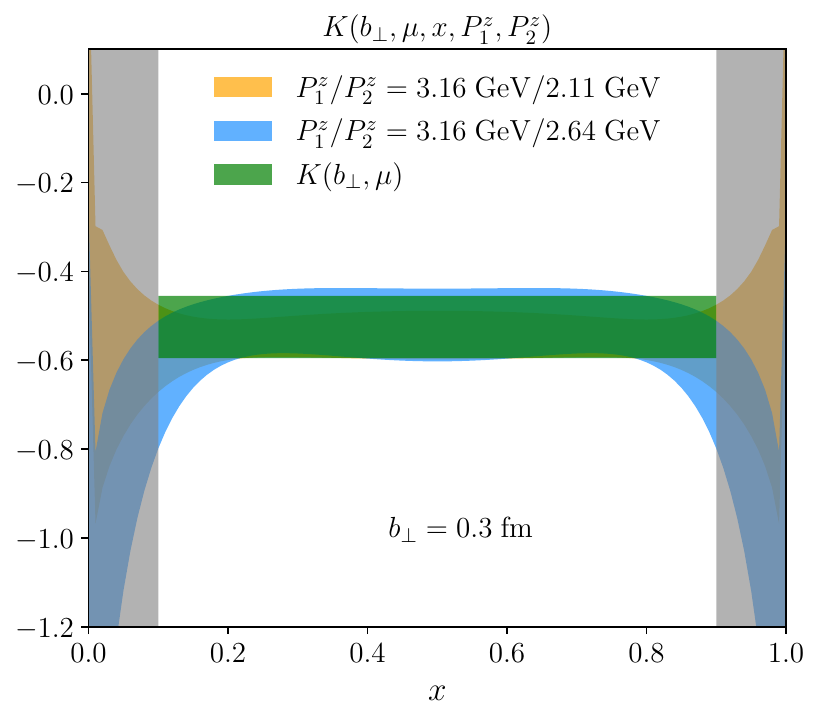}
\caption{The momentum-dependent and the momentum-independent fits for the CS kernel using the ansatz Eq.~(\ref{eq:cs-fit-model}).  We select results obtained at $b=0.3$fm for ensemble X650 as an example. Only the real parts and statistical uncertainties are shown. }
\label{fig:CS_fit}
\end{figure}

\begin{figure}[htb]
\centering
\includegraphics[width=0.75\textwidth]{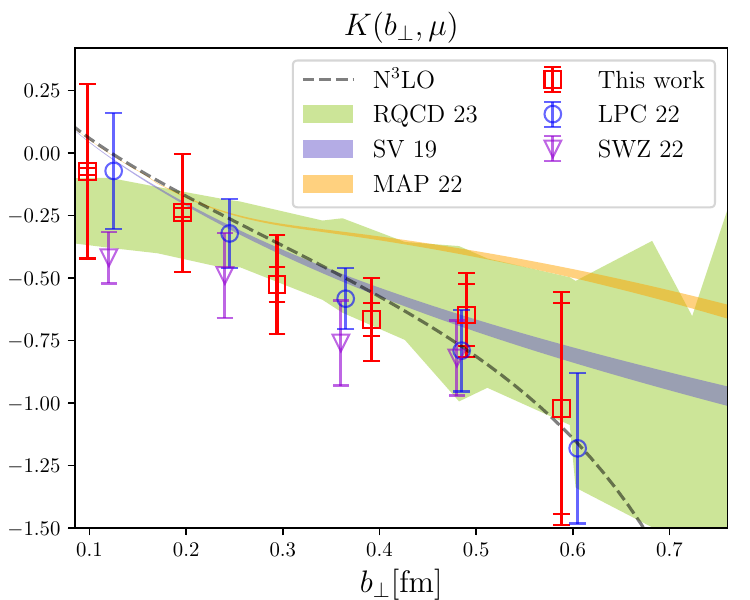}
\caption{A comparison of the CS kernel obtained in this work to the perturbative determination at three-loop order labelled as ``$\mathrm{N^3LO}$" \cite{ Li:2016ctv, Vladimirov:2016dll}, phenomenological extractions ``SV19" \cite{Scimemi:2019cmh} and ``MAP22" \cite{Bacchetta:2022awv}, and lattice calculations ``SWZ 22" \cite{Shanahan:2021tst}, ``LPC 22" \cite{LPC:2022ibr} and ``RQCD 23" \cite{Shu:2023cot}.}
\label{fig:CS_cls}
\end{figure}

In LaMET in principle both momenta should be large enough to significantly suppress the power corrections. For this reason we choose $P_1^z/P_2^z=3.16\ \mathrm{GeV}/2.11\ \mathrm{GeV}, 3.16\ \mathrm{GeV}/2.64\ \mathrm{GeV}$. To further extract the leading power contributions, namely to get rid of the finite momentum effects, we fit the momentum-dependent CS kernel to the following theoretically inspired ansatz \cite{LPC:2022ibr}
\begin{equation}
K(b_{\perp}, \mu, x, P_1^z, P_2^z)=K(b_{\perp}, \mu)+A\Big{[} \frac{1}{x^2(1-x)^2(P_1^z)^2} - \frac{1}{x^2(1-x)^2(P_2^z)^2} \Big{]}
\label{eq:cs-fit-model}
\end{equation}
in the range $x\in [0.1, 0.9]$. The intervals beyond this range are discarded as LaMET breaks down there. We show an example of this fit in Fig.~\ref{fig:CS_fit} for a selected $b_{\perp}$.  We point out that the CS kernel calculated in this way is complex and in Fig.~\ref{fig:CS_fit} only the real part is shown. The imaginary part comes from the matching kernel $H$, not the quasi TMDWF itself, see \cite{LPC:2022ibr}.  The final CS kernel result is shown in Fig.~\ref{fig:CS_cls} as red points. We take the real part as the central values. The statistical uncertainties are shown as inner error bars and the sum of the statistical and systematical uncertainties are shown as outer error bars. The systematical uncertainties are estimated using the measure
\begin{equation}
\sigma_{\mathrm{sys}}=\sqrt{\mathrm{Re}K(b_{\perp},\mu)^2+\mathrm{Im}K(b_{\perp},\mu)^2}-|\mathrm{Re}K(b_{\perp},\mu)|.
\end{equation}
From \cite{LPC:2022ibr} we know that the real part is equivalent to the average of the complex CS kernel calculated for both ``$\pm$" directions, which at the same time eliminates the imaginary part. 
 
 In Fig.~\ref{fig:CS_cls} we compare the CS kernel obtained in this work with those from other calculations, including the 3-loop perturbative results \cite{ Li:2016ctv, Vladimirov:2016dll}, the phenomenological extractions, SV19 \cite{Scimemi:2019cmh} and MAP22 \cite{Bacchetta:2022awv}, and the lattice calculations \cite{Shanahan:2021tst, LPC:2022ibr, Shu:2023cot}. The calculation \cite{Shanahan:2021tst} is based on the analysis of the quasi pion beam function with leading order matching kernel. The calculation \cite{LPC:2022ibr} is same as this work but on the MILC ensemble a12m310. The calculation \cite{Shu:2023cot} is based on the analysis of the (first) Mellin moments of the quasi TMDPDF, including one-loop contributions as well. It originally contains four data sets, obtained for pion and proton targets with twist-2 and twist-3 quasi TMDPDF operators. Here we have combined them and shown the results in a single band. The band is calculated by drawing Gaussian bootstrap samples at each $b_{\perp}$ value. Then the samples from different data sets are collected together, from which the median is taken as the expectation. The error is calculated by adding or subtracting the 34th percentiles on both sides of the median \cite{Altenkort:2023oms}. Note that in \cite{Shu:2023cot} only multiples of $b_{\perp}/a$ and square roots of sums of squares of $b_{\perp}/a$ have been considered, which explains the jagged shape of the band. We have interpolated between different $b_{\perp}$ linearly. From the comparison we can see that a general feature of the lattice determined CS kernel is that they suffer significant uncertainties. Within error the CS kernel obtained in this calculation is very close to the previous calculation performed on the MILC ensemble a12m310 using the same strategy \cite{LPC:2022ibr}, as expected. In addition these two are consistent with other lattice extractions within error. Not surprisingly they agree with the 3-loop perturbative results \cite{ Li:2016ctv, Vladimirov:2016dll} and the phenomenological extraction SV19 \cite{Scimemi:2019cmh} as well, in the small and moderate $b_{\perp}$ range. However all these results are lying below the recent phenomenological MAP22 fit \cite{Bacchetta:2022awv}, which is surprisingly flat. 
 
\begin{figure}[htb]
\centerline
{
\includegraphics[width=0.5\textwidth]{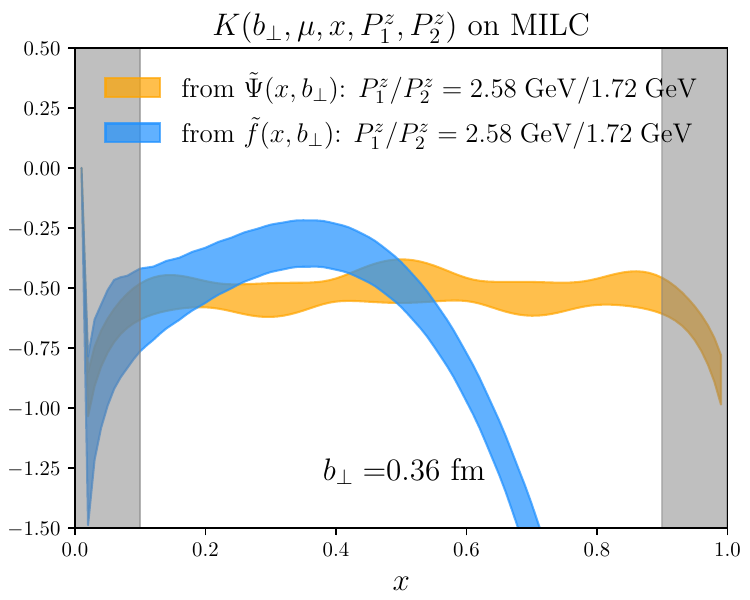}
\includegraphics[width=0.5\textwidth]{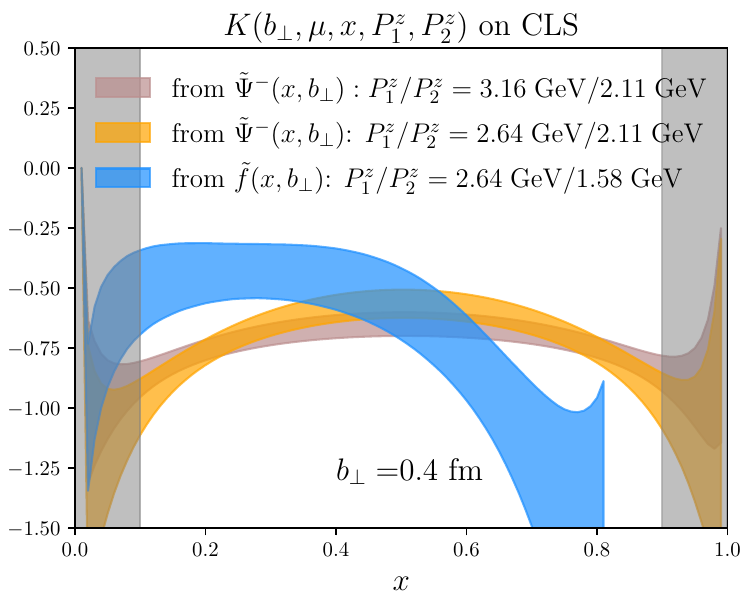}
}
\caption{The momentum-dependent CS kernel obtained from the quasi TMDWF and quasi TMDPDF on MILC ensembles (left) and CLS ensemble (right). In both panels the results are obtained from different ensembles: a12m130 for $\tilde{\Psi}$ and $\tilde{f}$ in the left panel, as well as X650 for $\tilde{\Psi}$ and A654 for $\tilde{f}$ in the right panel.}
\label{fig:CS_tmdpdf_tmdwf}
\end{figure}

In principle the CS kernel can also be obtained from the quasi TMDPDF via Eq.~(\ref{eq:CS-p-dep}), by replacing the quasi TMDWF objects with the quasi TMDPDF objects, and replacing the hard kernel function for quasi TMDWF with that for quasi TMDPDF. We have tried this and the results are shown in Fig. \ref{fig:CS_tmdpdf_tmdwf}. In the left panel we show the results obtained for the MILC ensemble a12m130 ($\tilde{f}$) and a12m310 ($\tilde{\Psi}$) at a moderate $b_{\perp}$ and in the right panel we show the results obtained for CLS ensemble A654 ($\tilde{f}$) and X650 ($\tilde{\Psi}$). In the left panel $\tilde{\Psi}=(\tilde{\Psi}^-+\tilde{\Psi}^+)/2$ so the resulting $K(b_{\perp}, \mu, x, P_1^z, P_2^z)$ is pure real while in the right panel only the real parts are shown because the data is lacking for $\tilde{\Psi}^+$. In the right panel $K(b_{\perp}, \mu, x, P_1^z, P_2^z)$ from $\tilde{\Psi}^-(x, b_{\perp})$ with momentum pair $2.64\ \mathrm{GeV}/2.11\ \mathrm{GeV}$ shows a poor plateau but this improves when moving to larger momentum pair $3.16\ \mathrm{GeV}/2.11\ \mathrm{GeV}$, consistent with the expectation from LaMET.
We want to stress that in this figure we only show the real part for the results obtained from quasi TMDWF while the results obtained from quasi TMDPDF are pure real by construction, which can be seen in Sec.\ref{sec:tmdpdf}. From this figure we find that for both ensembles the ratios from quasi TMDWF show better plateaus in $x$ while the ratios from quasi TMDPDF decay too fast after $x\sim 0.5$, which indicates an early breakdown of LaMET in this range. 
We remark that the difference we see between the results from quasi TMDWFs and quasi TMDPDFs is unlikely caused by the different pion mass or lattice volume. The common behavior we observed for both MILC and CLS ensembles suggests that this is a generic property.
Our tests show that using quasi TMDWF to extract the CS kernel will give better defined results, even though it suffers from systematic uncertainties induced by its imaginary part. To conclude, we will use the results obtained from quasi TMDWF (the squares and circles in Fig. \ref{fig:CS_cls}) as our final estimate for the CS kernel.

\section{Physical TMDs from the soft function}
\label{sec:app}
 With the intrinsic soft function and the CS kernel obtained in previous sections, we can extract the physical TMDs on the light cone based on the factorization Eq.~(\ref{eq:factorization_pdf}) or Eq.~(\ref{eq:factorization_wf}). We will consider first the  TMDWFs in Sec. \ref{sec:tmdwf} and then TMDPDFs in Sec. \ref{sec:tmdpdf}.
 
\subsection{TMDWFs in light cone}
\label{sec:tmdwf}

\begin{figure}[htb]
\centerline
{
\includegraphics[width=0.5\textwidth]{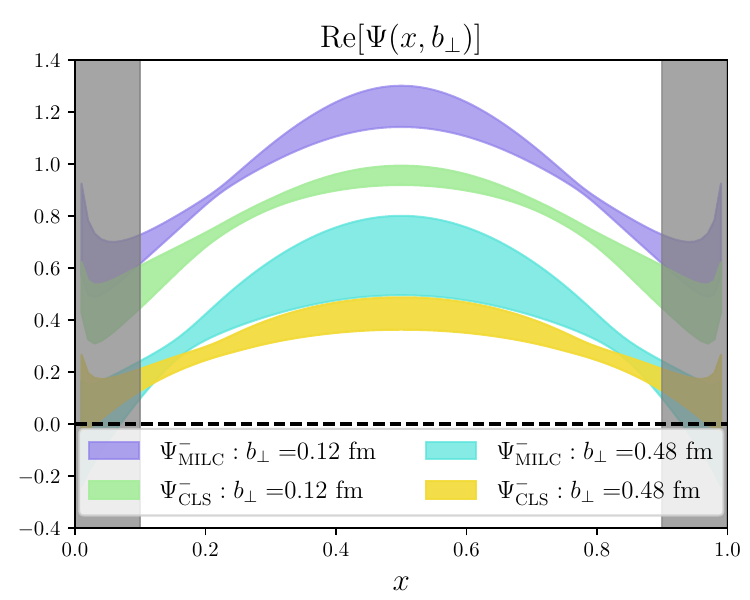}
\includegraphics[width=0.5\textwidth]{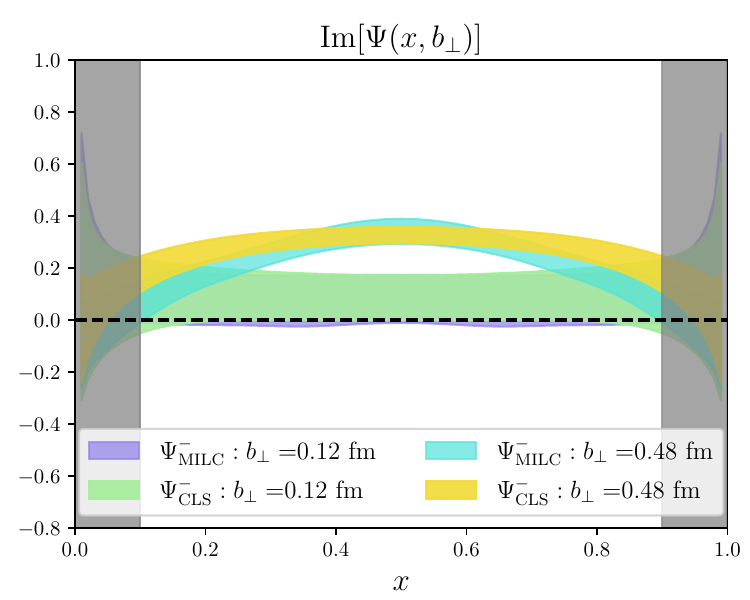}
}
\caption{The left panel shows the real parts of the physical TMDWFs from a12m310 and X650 while the right panel show the imaginary parts. They are determined at $\zeta=(6\;\mathrm{GeV})^2$, $\mu=2$~GeV and in the infinite $P^z$ limit. Only $\Psi^-$ is considered as an example.}
\label{fig:lcwf_mix}
\end{figure}

Inverting Eq.~(\ref{eq:factorization_wf}) one can obtain the physical TMDWFs. This is done for a12m310 ($a=0.121$ fm) and X650 ($a=0.098$ fm). The physical TMDWFs are extrapolated to infinite momentum using the ansatz Eq.~(\ref{eq:mom-extrap-soft}) with the intrinsic soft function replaced by the physical TMDWF. The matching was done at rapidity scale $\zeta=(6\;\mathrm{GeV})^2$ and renormalization scale $\mu=2$~GeV. There are various uncertainties contributing to the final physical TMDWFs. They are calculated in the same way as in \cite{Chu:2023jia}.

We show the physical TMDWFs in Fig.~\ref{fig:lcwf_mix}. The left panel shows the real parts 
%of the physical TMDWFs while 
and the right panel shows the imaginary parts. Note that as an example we only show the ``$-$" direction at a small and a large $b_{\perp}$. We remark that we have interpolated the X650 results using a cubic spline to the $b_{\perp}$ values analyzed for a12m310 to allow for a direct comparison. The top panels show that there exists visible discrepancies between X650 and a12m310 for the real parts at the two $b_{\perp}$ values (in fact, at intermediate $b_{\perp}$ values as well, which is not shown here) while for the imaginary part the tension is much reduced. A closer look at these discrepancies requires further investigation of the continuum limit.

Another finding is that the real parts of the amplitudes for both ensembles decrease with $b_{\perp}$, while for the imaginary parts it is the opposite. 

\subsection{TMDPDFs from the light cone}
\label{sec:tmdpdf}

Another application of the soft function is to determine the physical TMDPDFs. This has been done in \cite{LPC:2022zci} for a12m130 ($a=0.121$ fm), see Tab.\ref{Tab:setup}. In this work we show the results obtained for the CLS ensemble A654 ($a=0.098$ fm), aiming to understand the lattice discretization effects. The unpolarized bare quasi TMDPDF relevant for this calculations can be found in \cite{LPC:2022zci}. It should be mentioned that in this work we choose $\gamma^t$ in the bare quasi TMDPDF matrix element, as it approaches $\gamma^+$ at large momentum with smaller operator mixing effects \cite{LPC:2022zci}.

In the lattice simulations we put 20 sources on each configuration. $L$ is at most 10$a$ and $b_{\perp}/a$ is at most 7$a$. Both can be positive or negative. We average the two directions because they are equivalent. $z$ is at most 15$a$ and can also point into both directions. The real part of the three-point function is symmetric with respect to $z=0$ so it is averaged again. However the imaginary part is anti-symmetric so we need the multiply the negative part by $-1$ before averaging. In this way, we have 160 measurements per configuration. To control the excited states contamination, we calculate correlators at four different source-sink separations $t_{\mathrm{seq}}/a=\{5,6,7,8\}$ and extract the ground state contribution through a joint fit for all separations. The fits are always performed in the range $t\in[1, t_{\mathrm{seq}}-1]$. To examine the dependence on momentum, we consider three nucleon momenta $P^z\in \{3,4,5\}\times 2\pi/(24a)=\{1.58, 2.11, 2.64\}$ GeV. 

The bare quasi TMDPDF is renormalized in the same way as the quasi TMDWF using Wilson loop and $Z_O$ \cite{Zhang:2022xuw}. The difference is that $Z_O$ is now determined from the TMDPDF calculated in $\overline{\mathrm{MS}}$ scheme $\tilde{h}_{\Gamma}^{\mathrm{\overline{MS}}}\left(z_0, b_{\perp 0}, \mu\right)$, which turns out to be the same as the TMDWF in $\overline{\mathrm{MS}}$ scheme at zero momentum. On A654 the best plateau in $b_0$ appears at $z_0=0$, starting from $b_{\perp 0}=1a=0.1$ fm and we choose $z_0=0$, $b_{\perp 0}=1a$. With proper renormalization group evolution (RGE) as done in \cite{LPC:2022zci}, we find $Z_O=1.406(1)$. As for the large $L$ limit, following the spirit of Ref.~\cite{Zhang:2022xuw} and as verified above, we take the data at $L=7a=0.7$ fm at which the renormalized quasi TMDPDF has saturated to a constant.

For the $\lambda$ extrapolation, which uses the ansatz Eq.~(\ref{eq:lambda-ansatz}), we first fit the tails in the range $\lambda \in [9a, 12a]P^z$ at all momenta and estimate the central values and statistical uncertainties from this fit. Then we repeat the fit in the range $\lambda \in [7a, 10a]P^z$. The differences of the central values between the two fits are taken as the systematical uncertainties of the extrapolation. After $\lambda$-extrapolation we Fourier-transform the matrix elements obtained in position space to momentum space. We remark that the imaginary part vanishes due to its antisymmetry with respect to $z=0$. 

\begin{figure*}[htb]
\centerline{
\includegraphics[width=0.5\textwidth]{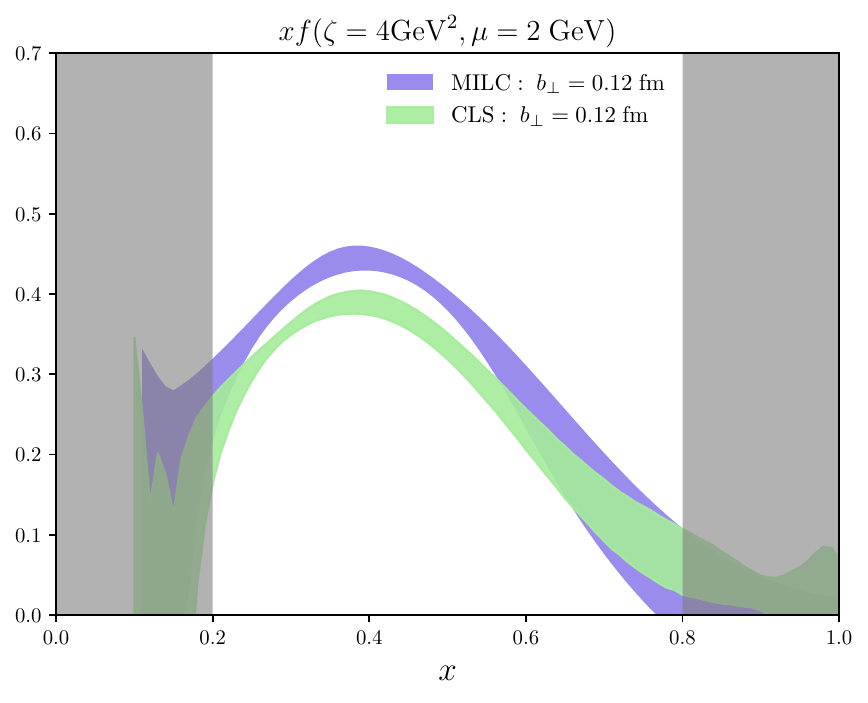}
\includegraphics[width=0.5\textwidth]{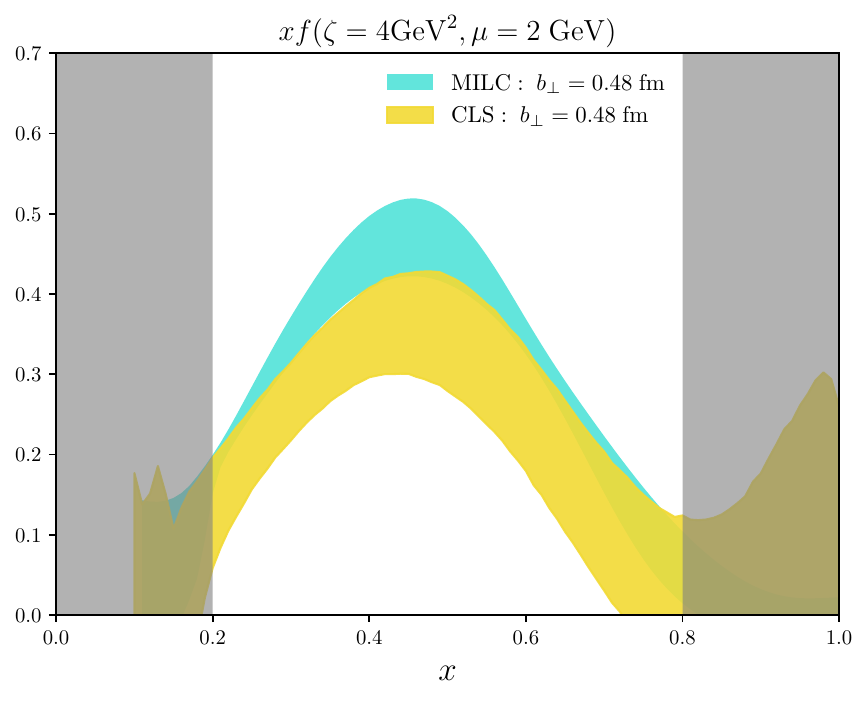}
}
    \caption{A comparison of the physical TMDPDFs obtained on CLS ensemble A654 and MILC ensemble a12m130 at $b_{\perp}$=0.12 fm (left) and 0.48 fm (right). On A654 we have interpolated TMDPDFs using a cubic spline to the $b_{\perp}$ values used for a12m130 for a direct comparison. }
    \label{fig:allb}
\end{figure*}

We now calculate the physical TMDPDF by applying the matching Eq.~(\ref{eq:factorization_pdf}) using the scale $\zeta=4\ \mathrm{GeV}^2$, followed by an extrapolation to infinite momentum. As in \cite{LPC:2022zci}, the RGE for the hard kernel Eq~(\ref{eq:hard_kernel_pdf}) is solved. In Appendix \ref{app:evolution} we illustrate the evolution of TMDWF and TMDPDF with respect to the renormalization scale $\mu$ and rapidity scale $\zeta$. 

In Fig.~\ref{fig:allb} we show the physical TMDPDFs obtained on A654 and compare them to those obtained on a12m130 \cite{LPC:2022zci}. The A654 results are again interpolated to the $b_{\perp}$ values attainable on a12m130. The endpoint regions $x<0.2$ and $x>0.8$ where LaMET breaks down have been grey shaded. Due to renormalization group resummation, the shaded regions are broader than in the case of the TMDWF. Overall speaking, the TMDPDFs on different ensembles have similar shape at the two $b_{\perp}$ values considered here, in fact at other $b_{\perp}$ as well, which we do not show; but given the uncertainties at $b_{\perp}=0.12$ fm and the observed discrepancies, further study of in particular discretization effects is needed to obtain reliable error estimates. 

\section{Conclusion}
\label{sec:summary}
We calculate the intrinsic soft function and the CS kernel on two different lattice ensembles. In this updated result of the soft function we have included the one-loop contributions and used proper normalization of light meson form factors. We have also suppressed higher-twist contaminations by Fierz rearrangements. We find that the intrinsic soft function obtained on two ensembles are similar except at small $b_{\perp}$ where lattice discretization effects are probably significant.

We have also tried to extract the CS kernel from TMDWFs including the new X650 ensemble in addition to those studied in \cite{LPC:2022ibr}. We also extract the CS kernel from quasi TMDPDFs. It turns out that using the former method is a better choice. We provide a comparison of the CS kernel obtained in this work and in other studies. We find that the CS kernel calculated on X650 shows good agreement with literature, particularly \cite{LPC:2022ibr}. Using the soft function we determine the physical TMDWFs for pion and physical TMDPDFs for proton from the corresponding quasi TMD objects renormalized using the method proposed in \cite{Zhang:2022xuw}, also on two ensembles. 
The good agreement observed on different ensembles for the TMDWFs and TMDPDFs verifies the applicability of calculating light cone quantities from lattice simulated quasi objects using TMD factorization via the soft function. From our findings we conclude that a determination of the soft function with better controlled precision and a systematic investigation of discretization effects is needed and possible. We leave this for future work.

%------------------------------------------------------------------------------------
%  acknowledgments
%------------------------------------------------------------------------------------

\section*{Acknowledgements}
We acknowledge the Rechenzentrum of Regensburg for providing computer time on the Athene Cluster. We thank the CLS Collaboration for sharing the ensembles used to perform this study. We thank Wolfgang S\"oldner for valuable discussions on the X650 ensemble. The LQCD calculations were performed using the multigrid algorithm~\cite{Babich:2010qb,Osborn:2010mb}, Chroma software suite~\cite{Edwards:2004sx} and QUDA~\cite{Clark:2009wm,Babich:2011np,Clark:2016rdz} through HIP programming model~\cite{Bi:2020wpt}. This work is supported in part by Natural Science Foundation of China under grant No. U2032102,  12125503, 12205106, 12175073, 12222503, 12293062, 12147140, 12205180. The computations in this paper were run on the Siyuan-1 cluster supported by the Center for High Performance Computing at Shanghai Jiao Tong University, and Advanced Computing East China Sub-center. J.H and J.L are also supported by Guangdong Major Project of Basic and Applied Basic Research No. 2020B0301030008, the Science and Technology Program of Guangzhou No. 2019050001.  Y.B.Y is also supported by the Strategic Priority Research Program of Chinese Academy of Sciences, Grant No. XDB34030303 and XDPB15. J.H.Z. is supported in part by National Natural Science Foundation of China under grant No. 11975051. J.Z. is also supported by Project funded by China Postdoctoral Science Foundation under Grant No. 2022M712088. A.S., H.T.S, W.W, Y.B.Y and J.H.Z are also supported by a NSFC-DFG joint grant under grant No. 12061131006 and SCHA~458/22.

%\clearpage

\section*{Appendix}
\appendix

\section{Dispersion relation}
\label{app:dispersion}
\begin{figure*}[thb]
\centerline
{
\includegraphics[width=0.5\textwidth]{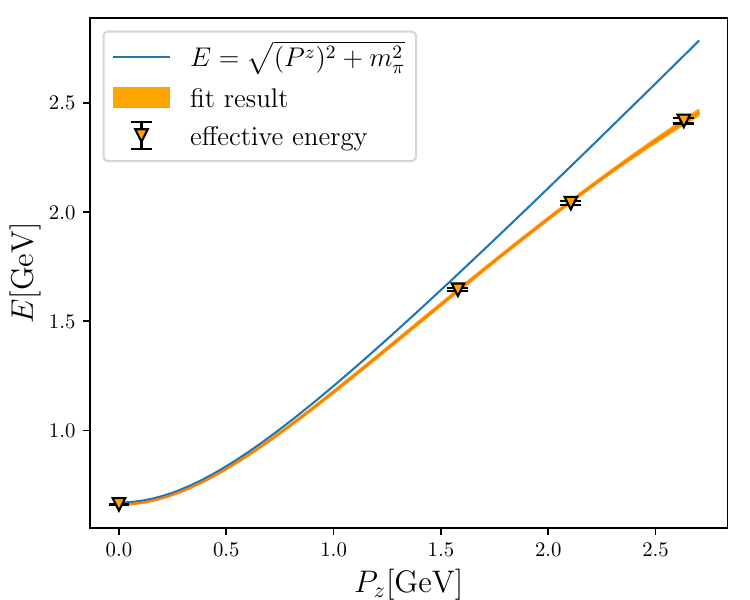}
\includegraphics[width=0.5\textwidth]{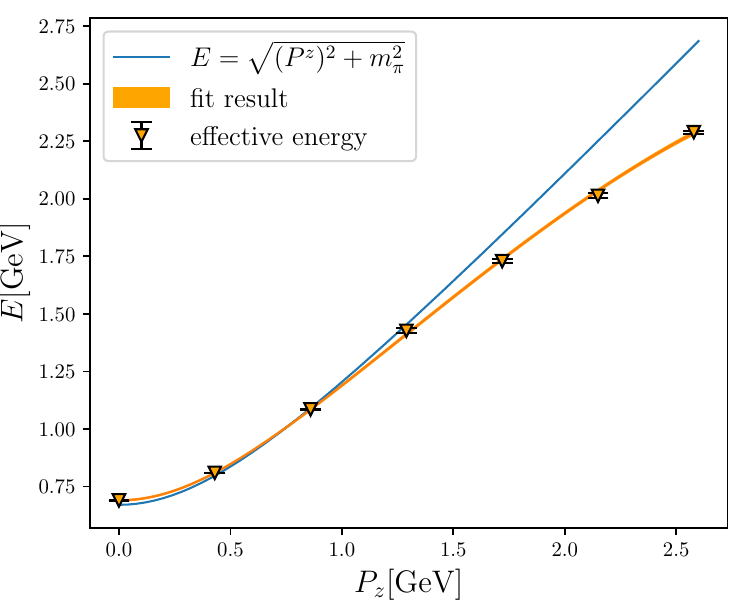}
}
\caption{Dispersion relation for the ensembles X650 (left) and a12m310 (right). The curves represent the fit of the lattice data with the ansatz $E_{\pi}=\sqrt{m^2_{\pi}+c_1 P^2 + c_2 P^4a^2}$. The fit values for $c_1$ and $c_2$ are 0.975(13), -0.116(11) for X650 and 0.9774(68), -0.1050(33) for a12m310. 
}
\label{fig:dispersion}
\end{figure*}

In this section we examine the dispersion relation on the ensembles X650 and a12m310, on which the soft functions are calculated. While the only small deviation of the lattice dispersion relation from the continuum one $E_{\pi}=\sqrt{m^2_{\pi}+P^2}$ suggests good control of lattice discretization effects, there is room for further improvement, which we will try to achieve in future work. The energies are extracted from the local two-point correlation functions of the pion. The results are summarized in Fig. \ref{fig:dispersion}. We have fit the extracted energies to the ansatz $E_{\pi}=\sqrt{m^2_{\pi}+c_1 P^2 + c_2 P^4a^2}$, in which the second term takes care of some of the discretization effects. Also the fit values of $c_2$ for both ensembles indicate noticeable but only moderate discretization effects.

\section{More examples for one-state fits for the ratio of the two-point correlation functions}
\label{app:t-fit}

\begin{figure*}[thb]
\centerline{
\includegraphics[width=0.5\textwidth]{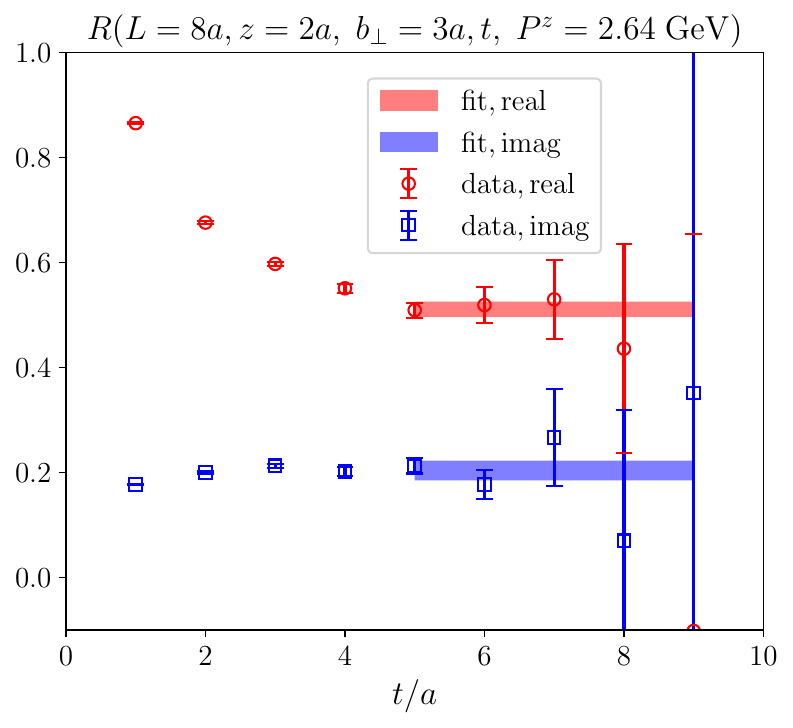}
\includegraphics[width=0.5\textwidth]{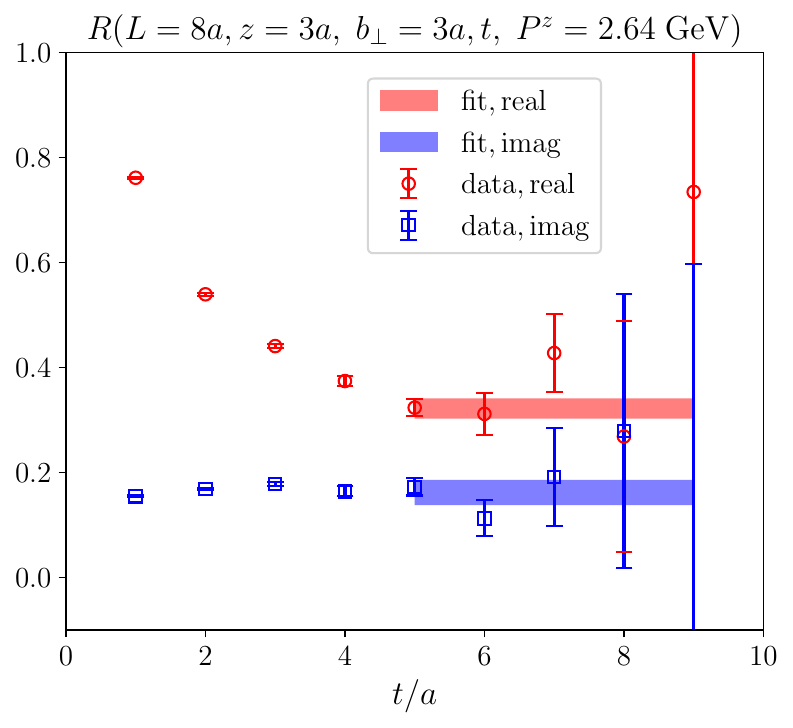}
}
\centerline{
\includegraphics[width=0.5\textwidth]{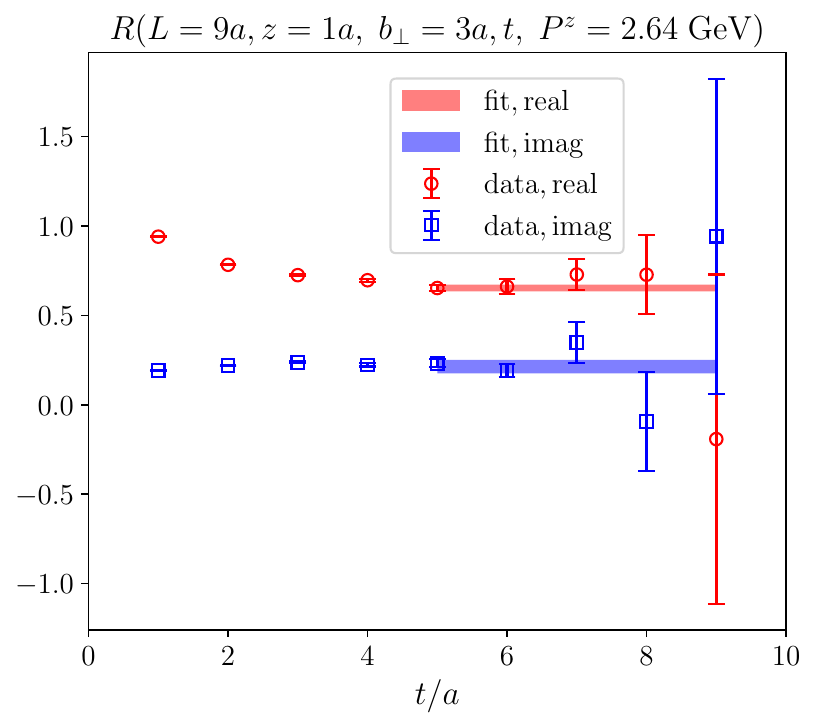}
\includegraphics[width=0.5\textwidth]{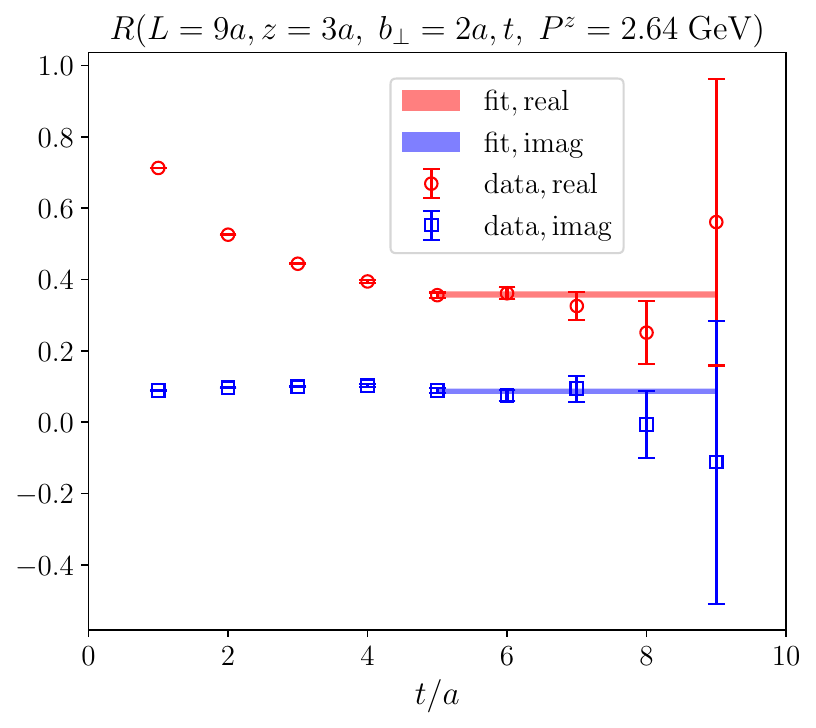}
}
\caption{More examples with $\{L,z,b_{\perp}\}$ for the one-state fit of Eq.~(\ref{eq:c2_fit}) at $P^z=2.64$ GeV.}
\label{fig:more-tfit}
\end{figure*}

In addition to the one-state fit of the ratio Eq.~(\ref{eq:c2_fit}) shown in Fig. \ref{fig:large_t_limit}, show in Fig. \ref{fig:more-tfit} some other examples to indicate that the results look always quite similar. In these cases we have a few typical sizes $\{L, z, b_{\perp}\}$ for the staple-shaped Wilson link. Based on this figure and Fig. \ref{fig:large_t_limit}, we consider our fit strategy as rational. 

\section{More examples of constant fits for large $L$}
\label{app:L-fit}
\begin{figure*}[thb]
\centerline
{
\includegraphics[width=0.5\textwidth]{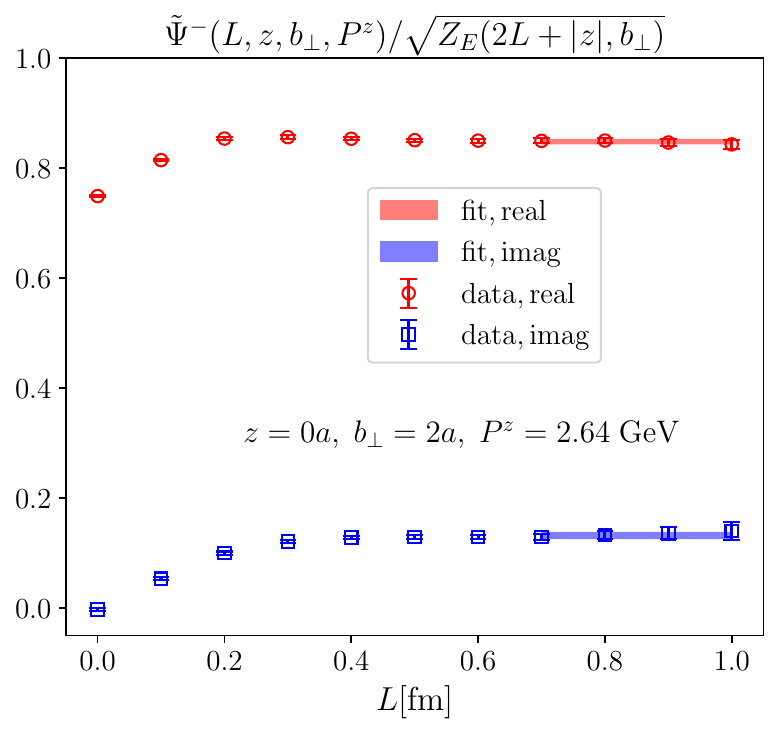}
\includegraphics[width=0.5\textwidth]{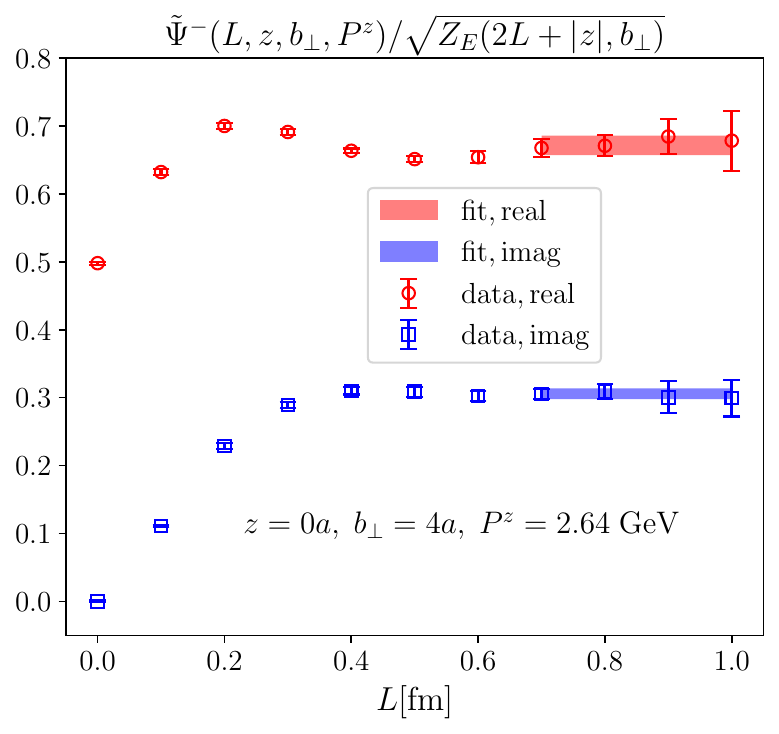}
}
\centerline
{
\includegraphics[width=0.5\textwidth]{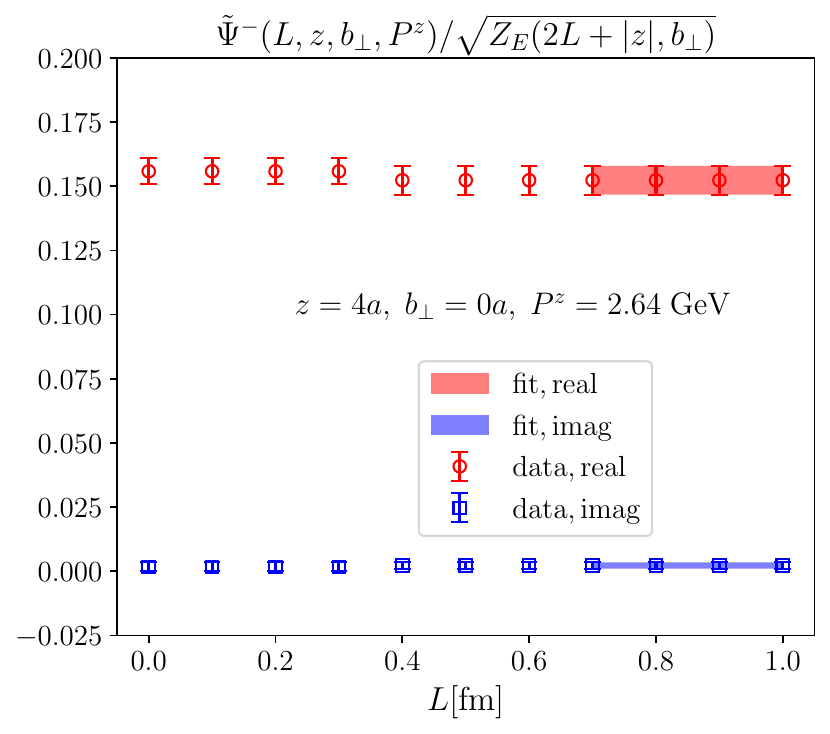}
\includegraphics[width=0.5\textwidth]{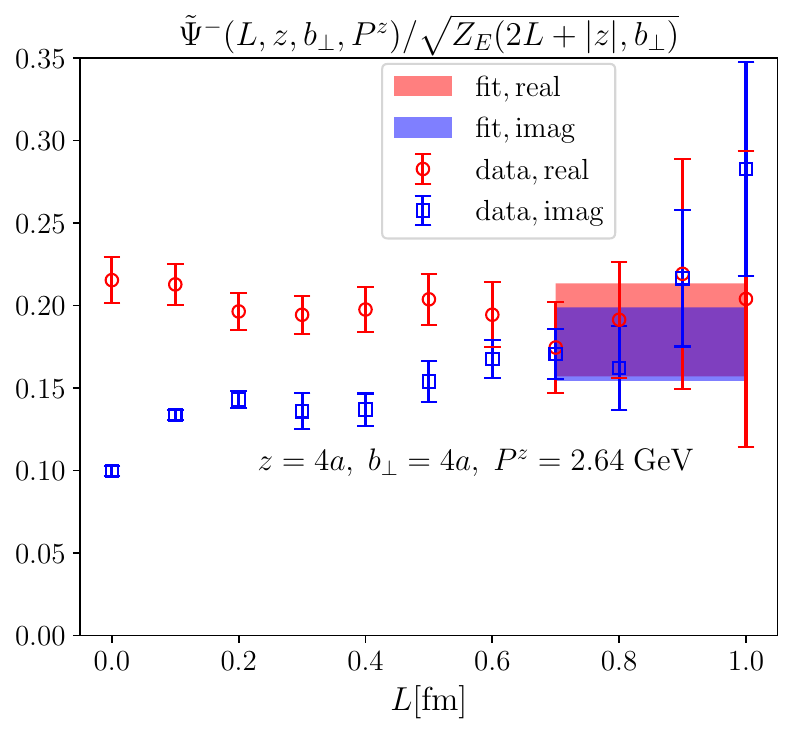}
}
\caption{More examples with $\{z,b_{\perp}\}$ for large $L$ fit at $P^z=2.64$ GeV.
}
\label{fig:more-large_L_limit}
\end{figure*}

Similar as in the above figures we show here some more for large $L$ fits. We consider in Fig. \ref{fig:more-large_L_limit} different $z$ and $b_{\perp}$ values. In all cases considered here, fitting to a constant at $L\geq 0.7$ fm is sufficient. Comparing the top panels and the bottom panels, one finds that for larger $z$, it is still safe to include more data points from smaller $L$ in the fit.

\section{Impact of sea quark mass to the determination of $S_I$}
\label{app:sea-quark-mass}

\begin{figure*}[thb]
\centerline{
\includegraphics[width=0.5\textwidth]{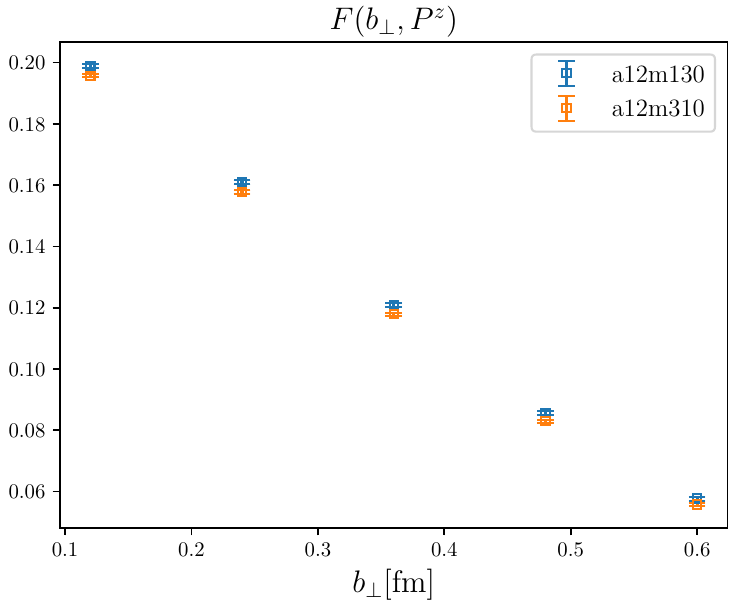}
\includegraphics[width=0.5\textwidth]{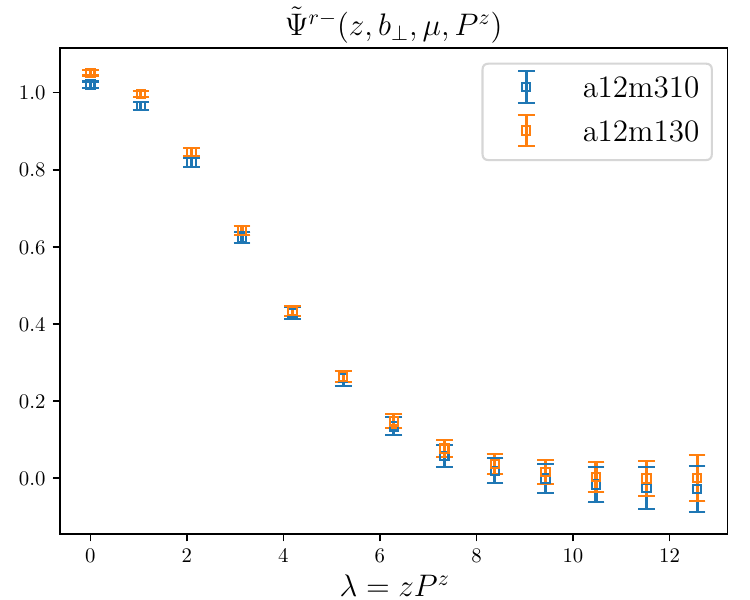}
}
\caption{Form factor (left) and quasi TMDWF (right) calculated at $P_z$=1.72 GeV on a12m130 and a12m310. In the calculation of the form factor we take the value at $t=t_{\mathrm{seq}}/2=4a$ as an estimate for the ground state contribution. In the right panel we choose a moderate $b_{\perp}$=0.36 fm.}
\label{fig:sea-quark-mass}
\end{figure*}

To examine the impact of sea quark mass on the determination of $S_I$, we calculate the form factor and quasi TMDWF at a selected momentum $P_z$=1.72 GeV on a12m130 and a12m310. For brevity, in the calculation of the form factor we consider only one single $t_{\mathrm{seq}}=8a$ and take the value at $t_{\mathrm{seq}}/2$ as an estimate for the $t_{\mathrm{seq}}\rightarrow\infty$ result, instead of performing a real fit. The results are shown in Fig. \ref{fig:sea-quark-mass}. It can be seen that for both quantities results from both ensembles give consistent results. The tiny difference is much smaller than the uncertainties from other aspects, e.g., large-$\lambda$ extrapolation and the matching procedure, and thus can be safely ignored.  

\section{Impact of small $t_{\mathrm{seq}}$ to the fit of three-point correlation function}
\label{app:small_tseq}

\begin{figure}[thb]
\centerline
{
\includegraphics[width=0.5\textwidth]{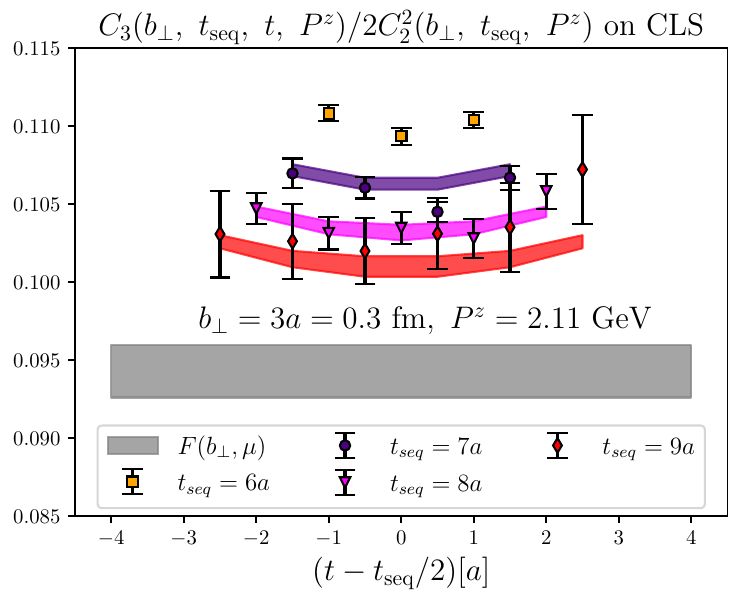}
}
\caption{Same as the right panel of Fig. \ref{fig:c3_c2_fit} but excluding $t_{\mathrm{seq}}=6a$ data set in the fit.
}
\label{fig:c3-fit-no-t6}
\end{figure}

To see how much the small $t_{\mathrm{seq}}$ data set affects the extrapolation to $t_{\mathrm{seq}}\rightarrow \infty$, we perform a fit of the same data set used in the right panel of Fig. \ref{fig:c3_c2_fit} but exclude $t_{\mathrm{seq}}=6a$. The results are shown in Fig. \ref{fig:c3-fit-no-t6}. Comparing to the right panel of Fig. \ref{fig:c3_c2_fit} it can be seen that only the error in $F(b_{\perp}, \mu)$ grows slightly while the change in the mean value lies within the statistical error. This confirms that the excited-state contamination is under control after the extrapolation. If we include the change coming from different data sets used in the extrapolation as systematic uncertainty, we can see it is of $\mathcal{O}(1\%)$, similar to the statistical uncertainty. Such change will be much smaller for the MILC ensemble where more $t_{\mathrm{seq}}$ are available. We stress that its influence is very limited in the sense of physics, when compared to, e.g., the change of matching from tree level to one-loop level, see Fig. \ref{fig:soft_function}. 

\section{Evolution of TMDWF and TMDPDF with $\zeta$ and $\mu$}
\label{app:evolution}

\begin{figure*}[thb]
\centerline{
\includegraphics[width=0.5\textwidth]{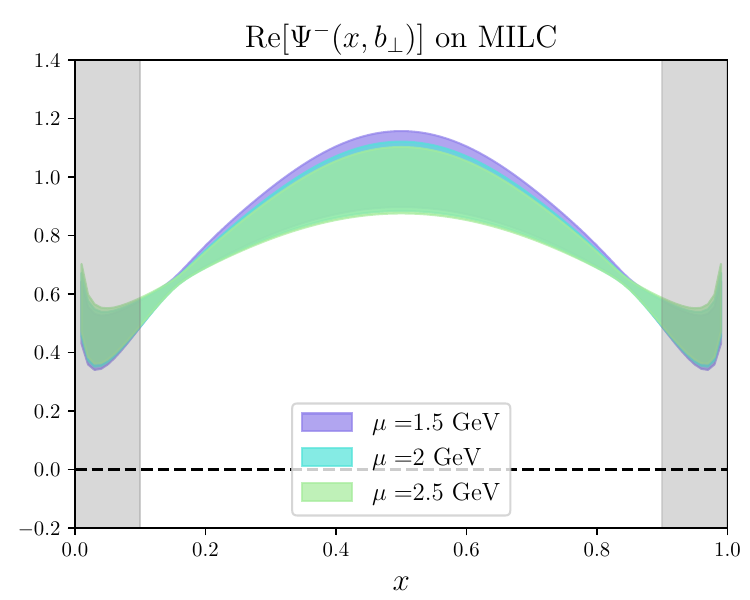}
\includegraphics[width=0.5\textwidth]{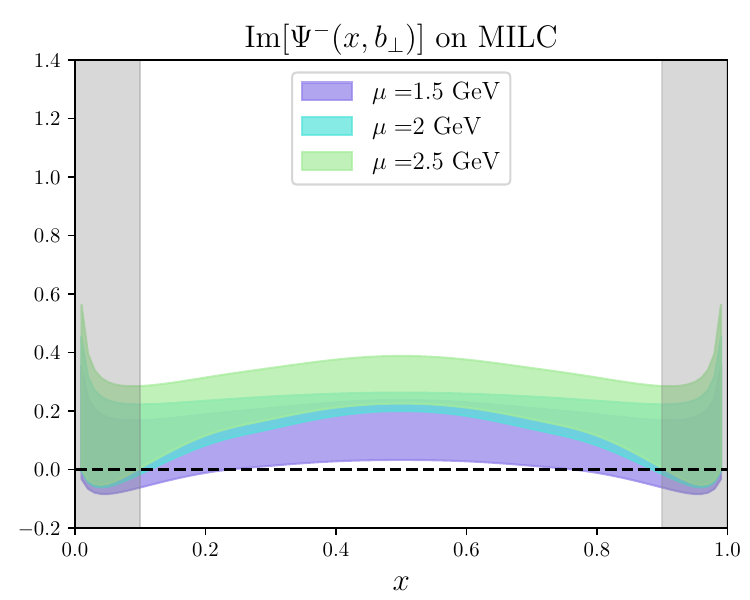}
}
\centerline{
\includegraphics[width=0.5\textwidth]{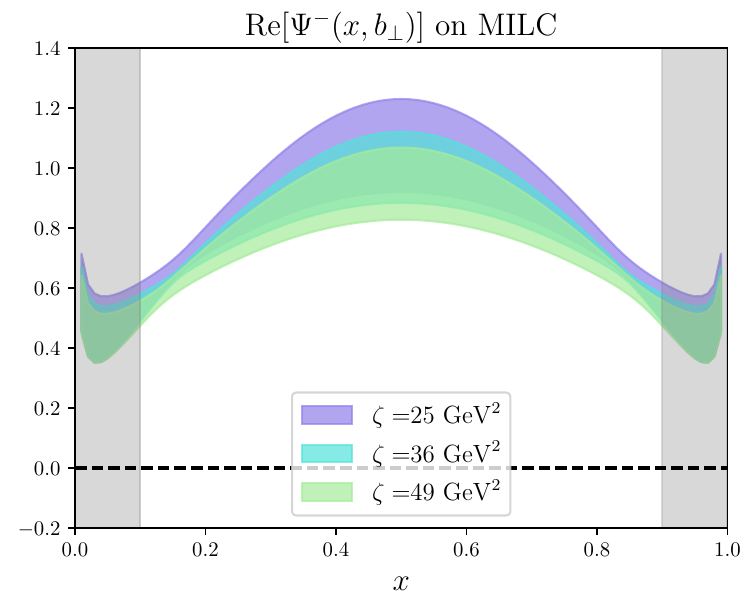}
\includegraphics[width=0.5\textwidth]{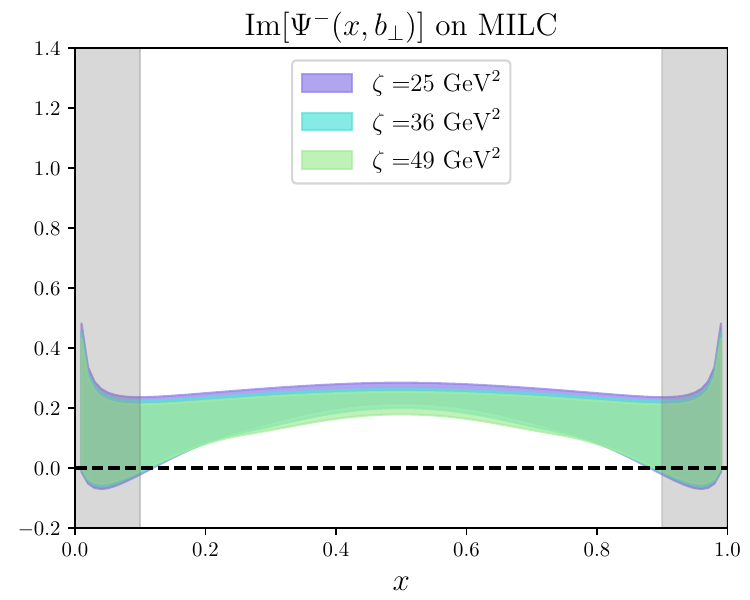}
}
\caption{The dependence of TMDWF on the renormalization scale $\mu$ (top) and the rapidity scale $\zeta$ (bottom).
}
\label{fig:evolution-tmdwf}
\end{figure*}

\begin{figure*}[thb]
\centerline{
\includegraphics[width=0.5\textwidth]{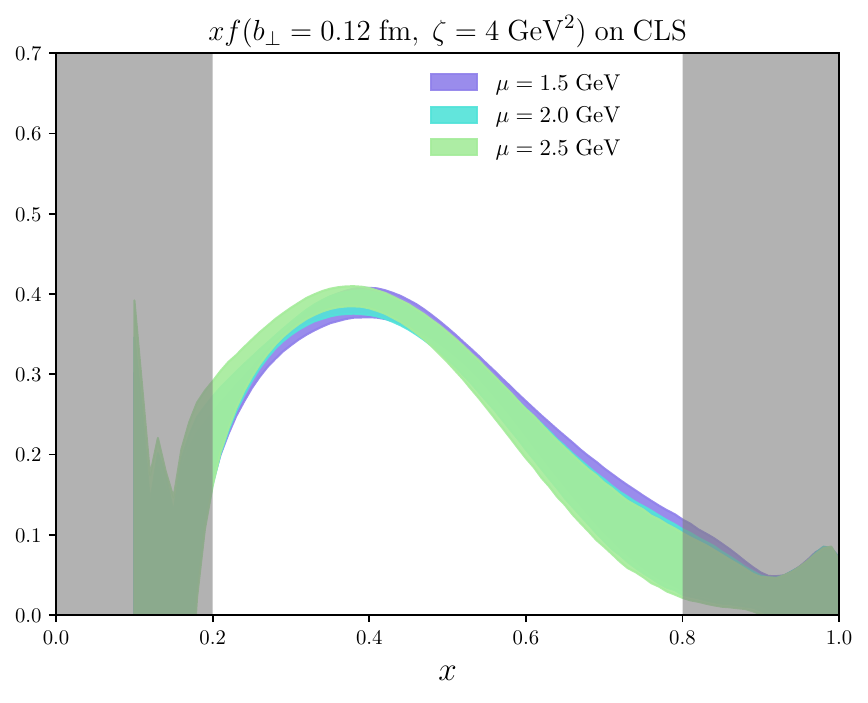}
\includegraphics[width=0.5\textwidth]{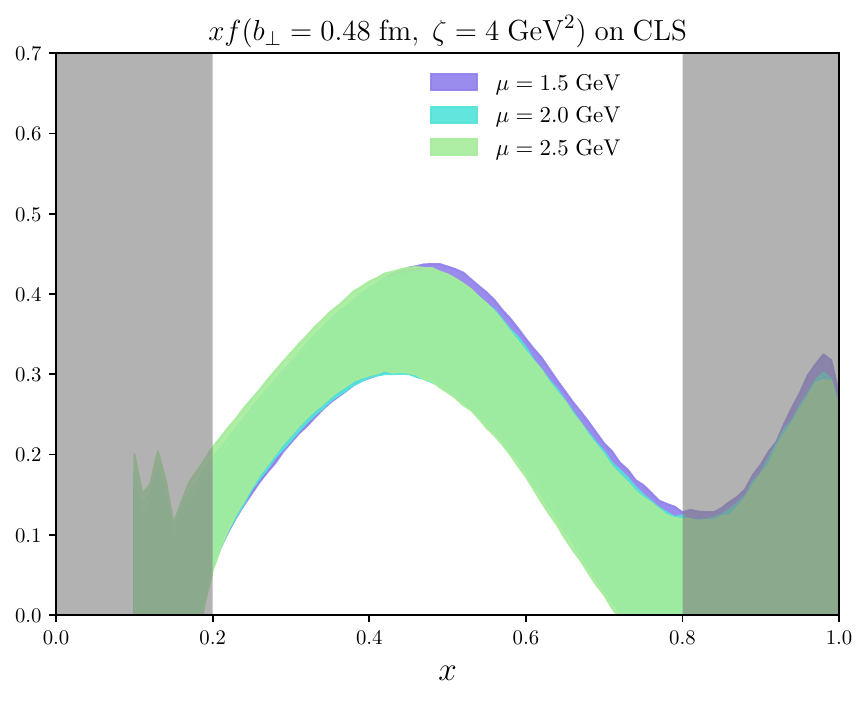}
}
\centerline{
\includegraphics[width=0.5\textwidth]{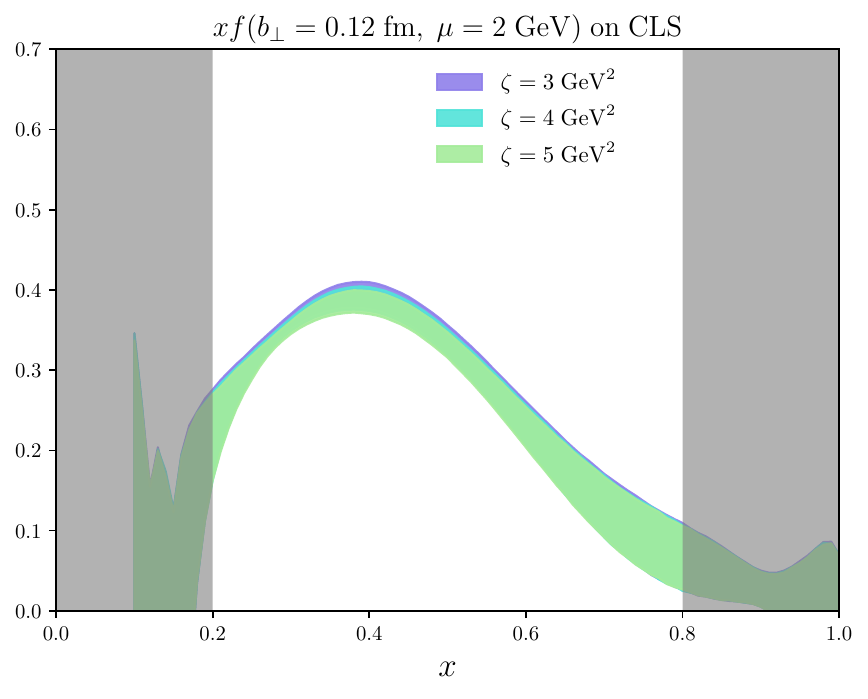}
\includegraphics[width=0.5\textwidth]{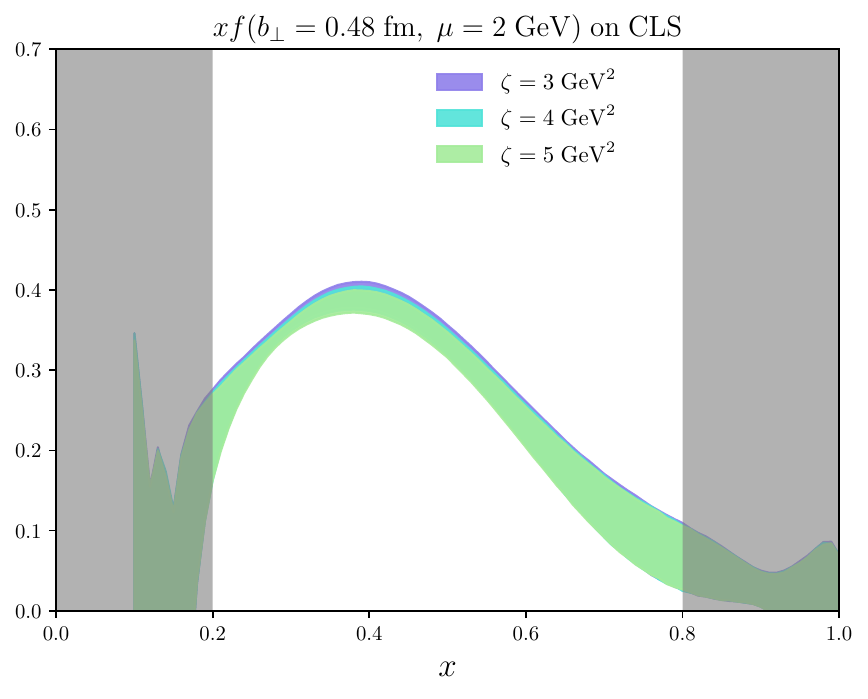}
}
\caption{The dependence of TMDPDF on the renormalization scale $\mu$ (top) and the rapidity scale $\zeta$ (bottom).
}
\label{fig:evolution-tmdpdf}
\end{figure*}

In this section we illustrate the evolution of TMDWF and TMDPDF with respect to the rapidity scale $\zeta$ and the renormalization scale $\mu$. For TMDWF we take a MILC ensemble as an example. We consider three renormalization scales $\mu=1.5, 2.0, 2.5$ GeV and three rapidity scales $\zeta=25, 36, 49$ GeV$^2$. The results are shown in Fig. \ref{fig:evolution-tmdwf}. For TMDPDF we take a CLS ensemble as an example. The renormalization scales are the same as above and the rapidity scales are $\zeta=3, 4, 5$ GeV$^2$. The results are shown in Fig. \ref{fig:evolution-tmdpdf}. It can be seen that in all cases TMDWF and TMDPDF have very mild dependence on both scales in the chosen range considered here.

%%------------------------------------------------------------------------------------
%       references
%------------------------------------------------------------------------------------
\bibliographystyle{JHEP}
\bibliography{paper}

\end{document}